\documentclass[11pt]{article}
\pdfoutput=1
\usepackage{amsfonts}
\usepackage{amsmath}
\usepackage{url}
\usepackage{hyperref}
\usepackage{slashed}
\usepackage{tikz}
\usepackage{graphicx}
\usepackage{epstopdf}
\usepackage{subfigure}
\usepackage{pgfplots}
\usepackage{amssymb}
\usepackage{color}
\usepackage{mathrsfs}
\usepackage{fancybox}
\usepackage{lipsum}
\usepackage{eurosym}
\usepackage{tcolorbox}

\def\({\left (}
\def\){\right )}
\def\[{\left [}
\def\]{\right ]}
\def\d{\mathrm{d}}

\numberwithin{equation}{section}
\def\Deltal{\ell}
\usepackage{environ}
\NewEnviron{myequation}{%
\begin{eqnarray}
\scalebox{1.05}{$\BODY$}
\end{eqnarray}
}

\newcommand{\beq}{\begin{equation}}
\newcommand{\eeq}{\end{equation}}
\newcommand{\nn}{\nonumber\\} 
\newcommand{\bea}{\begin{eqnarray}}
\newcommand{\ea}{\end{eqnarray}}
\newcommand{\barr}{\!\begin{array}}
\newcommand{\earr}{\end{array}\!}
\newcommand{\lb}{{\langle}}
\newcommand{\rb}{{\rangle}}

\def\d{{\partial}}
\def\n{{\bf \widehat n}}
\def\k{{\bf k}}

\usepackage{mciteplus}

\begin{document}
\begin{titlepage}

\setcounter{page}{1} \baselineskip=15.5pt \thispagestyle{empty}

\vfil

${}$
\vspace{1cm}

\begin{center}

\def\thefootnote{\fnsymbol{footnote}}
\begin{changemargin}{0.05cm}{0.05cm} 
\begin{center}
{\Large \bf Shockwave S-matrix from Schwarzian Quantum Mechanics}
\end{center} 
\end{changemargin}

~\\[1cm]
{Ho Tat Lam${}^{\rm a}$\footnote{\href{mailto:htlam@princeton.edu}{\protect\path{htlam@princeton.edu}}}, Thomas G. Mertens${}^{\rm a,c}$\footnote{\href{mailto:thomas.mertens@ugent.be}{\protect\path{thomas.mertens@ugent.be}}}, Gustavo J. Turiaci${}^{\rm a}$\footnote{\href{mailto:joaquinturiaci@gmail.com}{\protect\path{joaquinturiaci@gmail.com}}} and Herman Verlinde${}^{\rm a,b}$\footnote{\href{mailto:verlinde@princeton.edu}{\protect\path{verlinde@princeton.edu}}}}
\\[0.3cm]

{\normalsize { \sl ${}^{\rm a}$Physics Department and ${}^{\rm b}$Princeton Center for Theoretical Science 
\\[1.0mm]
Princeton University, Princeton, NJ 08544, USA}} \\[3mm]
{\normalsize { \sl ${}^{\rm c}$Department of Physics and Astronomy
\\[1.0mm]
Ghent University, Krijgslaan, 281-S9, 9000 Gent, Belgium}}

\end{center}


 \vspace{0.2cm}
\begin{changemargin}{01cm}{1cm} 
{\small  \noindent 
\begin{center} 
\textbf{Abstract}
\end{center} }
Schwarzian quantum mechanics describes the collective IR mode of the SYK model and captures key features of 2D black hole dynamics. Exact results for its correlation functions were obtained in JHEP {\bf 1708}, 136 (2017) [arXiv:1705.08408]. We compare these results with bulk gravity expectations. We find that the semi-classical limit of the OTO four-point function exactly matches with the scattering amplitude obtained from the Dray-'t Hooft shockwave $\mathcal{S}$-matrix.  We show that the two point function of heavy operators reduces to the semi-classical saddle-point of the Schwarzian action. We also explain a previously noted match between the OTO four point functions and 2D conformal blocks. Generalizations to higher-point functions are discussed.

\end{changemargin}
 \vspace{0.3cm}
\vfil
\begin{flushleft}
\today
\end{flushleft}

\end{titlepage}

\newpage
\tableofcontents
\newpage

\addtolength{\abovedisplayskip}{.5mm}
\addtolength{\belowdisplayskip}{.5mm}

\def\plus{\raisebox{.5pt}{\tiny$+$\smpc}}

\addtolength{\parskip}{.6mm}
\def\spc{\hspace{1pt}}

\def\nspc{{\hspace{-2pt}}}
\def\ff{\rm\smpc f\smpc} 
\def\fff{\mbox{Y}}
\def\ww{{\rm w}}
\def\smpc{{\hspace{.5pt}}}

\def\zz{{\spc \rm z}}
\def\xx{{\rm x\smpc}}
\def\xxi{\mbox{\footnotesize \spc $\xi$}}
\def\jj{{\rm j}}

\renewcommand{\Large}{\large}

\def\calO{{b}}
\def\be{\begin{equation}}
\def\ee{\end{equation}}
\def\mathbi#1{\textbf{\em #1}} 
\def\som{{ \textit{\textbf s}}} 
\def\tom{{ \textit{\textbf t}}} 
\def\la{\langle}
\def\bea{\begin{eqnarray}}
\def\eea{\end{eqnarray}}
\def\is{\!  & \!  = \!  &  \!}
\def\ra{\rangle}
\def\half{{\textstyle{\frac 12}}}
\def\cL{{\cal L}}
\def\halfi{{\textstyle{\frac i 2}}}
\def\ba{\bea}
\def\ea{\eea}
\def\lb{\langle}
\def\rb{\rangle}
\newcommand{\rep}[1]{\mathbf{#1}}

\def\uU{\bf U}
\def\be{\bea}
\def\ee{\eea}
\def\delbar{\overline{\partial}}
\def\ra{\bigr\rangle}
\def\la{\bigl\langle}
\def\ccdot{\!\spc\cdot\!\spc}
\def\nspc{\!\spc\smpc}
\def\tr{{\rm tr}}
\def\ra{\bigr\rangle}
\def\la{\bigl\langle}
\def\li{\bigl|\spc}
\def\ri{\bigr |\spc}

\def\hf{\textstyle \frac 1 2}

\def\bfcdot{\raisebox{-1.5pt}{\bf \LARGE $\spc \cdot\spc $}}
\def\spc{\hspace{1pt}}
\def\is{\! &  \! = \! & \!}
\def\d{{\partial}}
\def\n{{\bf \widehat n}}
\def\k{{\bf k}}
\def\GO{{\cal O}}

\def\pp{{\mbox{\tiny$+$}}}
\def\mm{{\mbox{\tiny$-$}}}

\setcounter{tocdepth}{2}
\addtolength{\baselineskip}{0.2mm}
\addtolength{\parskip}{.6mm}
\addtolength{\abovedisplayskip}{.5mm}
\addtolength{\belowdisplayskip}{.5mm}

\def\fff{e}

\setcounter{footnote}{0}

\section{Introduction}\label{sec:Sch}
\vspace{-2.5mm}

Black holes induce gravitational shockwave interactions between infalling and outgoing particles near the horizon \cite{Dray:1984ha}. When reduced to 1+1 dimension, the resulting scattering matrix takes the following simple form \cite{Dray:1984ha}
\bea
\label{thds}
\mathcal{S} \is \exp{\bigl(i \kappa \hspace{1mm} p_- p_+\bigr)}.
\eea
Here $p_+$ and $p_-$ denote the Kruskal momentum operators of the incoming and outgoing particle, and $\kappa$ is the Newton constant.
Equation \eqref{thds} defines a manifestly unitary 2-to-2 scattering process. It reflects the geometric statement that when two highly boosted particles collide, the final Kruskal positions $U$ and $V$ of the particles are related to the initial positions via a simple coordinate shift proportional to the Kruskal momentum of the other particle
\bea
\label{ushift}
U \to  U \, + \, \kappa\, p_- ,\quad &&\quad V \to V \, + \, \kappa \, p_+ .
\eea
This description of the gravitational scattering becomes accurate in the region very close to the horizon,  located at $V=0$ and $U=0$. In this region, the Kruskal coordinates are related to the Schwarzschild coordinates via $U=-e^{2\pi u/\beta}$ and $V=e^{-2\pi v/\beta}$ where $\beta$ is the inverse temperature of the black hole. The shockwave interaction thus represents an exponentially growing effect in the Schwarzschild coordinate frame.

This shockwave interaction was shown to lead to exponentially growing commutators between infalling and outgoing modes in \cite{Kiem:1995iy}. In recent years, it was recognized that in holographic settings this exponential growth is a manifestation of maximally chaotic quantum dynamics of the underlying microscopic theory, and can be exhibited by studying a suitable class of out-of-time ordered correlation functions \cite{SS, JMV, Shenker:2014cwa,KitaevTalks, MSS}.

An interesting class of solvable theories that displays maximally chaotic behavior are the Sachdev-Ye-Kitaev (SYK) models \cite{KitaevTalks, Sachdev:1992fk, Polchinski:2016xgd, Maldacena:2016hyu, Jevicki:2016bwu, Cotler:2016fpe}. As first recognized by Kitaev \cite{KitaevTalks} (see also \cite{KS}), the IR dynamics of the SYK model is dominated by a single effective degree of freedom $f(\tau)$ representing reparametrizations of a 1D circle (throughout the paper $\tau$ will label Euclidean time, while $t$ will label Lorentzian time), with an unusual action that consists of the Schwarzian derivative
\bea
\label{schwaction}
\qquad S[f]
\is  - C\int_0^{\beta}\!\!\! d\tau \spc \bigl\{\spc F,\spc \tau\spc \bigr\}, \quad\qquad F\, \equiv \, \tan\left(\frac{\pi f}{\beta}\right),
\eea
where $\bigl\{\spc F,\spc \tau\spc \bigr\} = \frac {F'''}{F'} - \frac 3 2 \bigl(\frac{F''}{F'}\bigr)^2. \,$ The variable $f(\tau+\beta) = f(\tau) +\beta$  defines an element of the group Diff$(S^1)$ of diffeomorphisms of the thermal circle. The parameter $C$ is a dimensionful constant. 
This action is also found to describe 2D Jackiw-Teitelboim dilaton gravity with suitable asymptotic boundary conditions \cite{Almheiri:2014cka, Maldacena:2016upp, Engelsoy:2016xyb, Cvetic:2016eiv, Nayak:2018qej}. 

\def\TO{{\rm TO}}
\def\OTO{{\rm OTO}} 
In \cite{MTV}, the Schwarzian theory was shown to arise as a suitable limit of 2D Virasoro conformal field theory.\footnote{The role of the Schwarzian theory relative to the SYK model is indeed similar to that of Liouville theory relative to any holographic 2D CFT \cite{JMV, Turiaci:2016cvo, Turiaci:2017zwd}. Both theories capture the dynamics of geometric effective IR degrees of freedom and are manifestly linked with AdS gravity in one higher dimension. Both are also exactly solvable.}
This relation was then used to obtain exact expressions for its correlation functions, see also \cite{altland, Cotler:2016fpe, StWi, Zhenbin} for a different approach. 
In this paper we will focus on the out-of-time ordered (OTO) four-point function $\langle V_1 W_3 V_2 W_4\rangle$
(with $V_1 = V(t_1)$, etc) at finite inverse temperature $\beta$. The answer for the four point function can be written in the form of a momentum space integral
\bea \label{intro:OTOC}
\la V_1 W_3 V_2 W_4\ra \is \prod_{i=1,4,s,t}\int d k_i^2\sinh 2\pi k_i\  {\cal A}_{\rm OTO}(k_i,t_i)
\eea
where $k_i$ labels the energy of the intermediate states via $E_i = {k_i^2}/{2C}$ and $d k^2 = 2 k dk$. The explicit form of the momentum space amplitudes ${\cal A}_{\rm OTO}(k_i,t_i)$ is given in section 3.2. It can be diagrammatically represented as
$$
\begin{tikzpicture}[scale=.65, baseline={([yshift=0cm]current bounding box.center)}]
\draw[thick] (-1.05,1.05) -- (-.15,.15);
\draw[thick] (.15,-.15) -- (1.05,-1.05);
\draw[thick] (-1.05,-1.05) -- (1.05,1.05);
\draw[thick] (0,0) circle (1.5);
\draw[fill,black] (-1.05,-1.05) circle (0.08);
\draw[fill,black] (1.05,-1.05) circle (0.08);
\draw[fill,black] (-1.05,1.05) circle (0.08);
\draw[fill,black] (1.05,1.05) circle (0.08);
\draw(-3.7,0) node {${\cal A}_\OTO
\, =$};
\draw (1.85,0) node {\footnotesize $k_s$};
\draw (-1.85,0) node {\footnotesize $k_t$};
\draw (-.75,.33) node {\footnotesize $\ell_2$};
\draw (.78,.33) node {\footnotesize $\ell_1$};
\draw (0,1.75) node {\footnotesize $k_1$};
\draw (0,-1.75) node {\footnotesize $k_4$};
\draw (-1.4,-1.2) node {\footnotesize $t_2$};
\draw (-1.4,1.2) node {\footnotesize $t_3$};
\draw (1.4,-1.2) node {\footnotesize $t_4$};
\draw (1.4,1.2) node {\footnotesize $t_1$};
\end{tikzpicture}\qquad
$$
Here the lines connect the identical pairs of operators, each placed at different times along the thermal circle. The OTO property means that, in contrast with the geometric ordering, the time instances are ordered via $t_1 < t_2 < t_3 < t_4.$ The physical properties of the OTO amplitude, including Lyapunov growth, can be deduced equally well from position or momentum space.

The OTO four point function encodes direct information about the chaotic behavior of the quantum theory and about the gravitational scattering in the bulk dual. Indeed, we can think of the two lines in the above diagrams as world lines of two bulk particles.  In the OTO case, the two worldlines cross, indicating that the amplitude contains a non-trivial factor in the form of an $R$-matrix. This R-matrix captures the gravitational shockwave scattering in the bulk.

The shockwave interaction plays a key role in the Gao-Jafferis-Wall protocol \cite{GJW} for sending a signal through a wormhole of an eternal black hole. In \cite{MSY},  this protocol was refined and tested via the proposed identification between the thermo-field double state of the SYK model and the two-sided AdS${}_2$ black hole geometry. An important motivation for our study is to see whether the exact results for the correlation functions can be used to give further support for this proposal. We will focus on the large $C$ limit, or equivalently, the high temperature regime. In the bulk, this corresponds to the kinematic regime very close to the black hole horizon.

In this paper we study the large $C$ limit of the exact results \cite{MTV} and compare with semi-classical bulk calculations. The bulk answer for the 2-to-2 scattering amplitude due to the geometric shockwave interaction takes the form of an integral expression
\bea
\label{shockoverlap}
\lb V_1 W_3 V_2 W_4\rb \is \int^{\infty}_0\! \frac{dp_+}{p_+} \int^{\infty}_0\! \frac{dp_-}{p_-} \; \Psi_{1}^*(p_+)\, \Phi^*_{3}(p_-)\; \mathcal{S}(p_-, p_+) \; \Psi_{2}(p_+)\,
\Phi_{4}(p_-).
\eea
where $ \mathcal{S}(p_-,p_+) = \exp\( \frac {i\beta} {4\pi C} p_+p_-\) $ is the 2D Dray-'t Hooft $ \mathcal{S} $-matrix \cite{Dray:1984ha} and $\Psi_1(p_+) = \Psi(t_1,p_+)$, etc, are suitable asymptotic wavefunctions with given Kruskal momentum (more details given in section \ref{sect4} below). In section \ref{sect5}, we show that our exact expressions for the OTO four point functions at large $C$ precisely reduces to this semi-classical result. This confirms our claim that the  formulas \eqref{intro:OTOC} and \eqref{exactoto} contain the geometric shockwave S-matrix as a identifiable subfactor. 

This paper is organized as follows.
In section \ref{sect2} we compute the matrix elements of  the Dray-'t Hooft S-matrix and discuss its relevance to wormhole traversability.
 In section \ref{sect3}  we summarize the exact solution of the Schwarzian. Section \ref{sect4} discusses the asymptotic wave functions and the two-point function of the Schwarzian theory. In section \ref{sect5} we
demonstrate that the exact OTO four-point function of the Schwarzian theory at large  $C$ exactly matches with the shockwave scattering amplitude (\ref{shockoverlap}).  We also explain why the shockwave amplitude (\ref{shockoverlap}) coincides with a degenerate limit of a  Virasoro conformal block.   Finally, in section \ref{sectell} we take the semi-classical large mass limit of two-point correlators, where the mass also scales with $C$. The resulting expressions are compared to solutions of the Schwarzian equations of motion, and provide non-trivial checks on the exact formulas of \cite{MTV}. Section \ref{conclusion} contains some concluding comments. Some technical results are explained in the appendices.


\section{The 2D Shockwave S-matrix}
\label{sect2}
\vspace{-2.5mm}

While seemingly perfectly causal, the shockwave interaction is intrinsically non-local. In this section we will make this non-locality more explicit by decomposing the scattering matrix \eqref{thds} in a Schwarzschild energy eigenbasis. We will first do the computation in a first quantized setting and then transfer the results to the second quantized field theory.

\vspace{-1mm}

\subsection{First quantized S-matrix}

\vspace{-2mm}

For a given Schwarzschild energy $\nu$, there are four types of modes: left infalling, right infalling, left outgoing and right outgoing. They are, up to normalizations
\bea
\label{lrmodes}
\la U \bigl| \psi_{{\!}_{\rm L,R}} ( \nu)\ra\! \is \! |U|^{\mp i\nu-1/2} \, \theta(\pm U)\qquad ; \qquad \la V \li  \phi_{{\!}_{\rm L,R}} ( \nu)\ra
\, = \, |V|^{\mp i\nu-1/2} \, \theta(\pm V)\, .
\eea
Alternatively, we can choose to distinguish four types of modes according to the direction and sign of their Kruskal momenta
\bea
\label{pmmodes}
\la p_+ \bigl| \psi_\pm(\nu)\ra \! \is \! |p_+|^{-i\nu-1/2} \, \theta(\pm\spc p_+)\qquad ; \qquad \la p_-\li  \phi_\pm(\nu) \ra \, = \, |p_-|^{ i\nu-1/2} \, \theta(\pm\spc p_-)\, ,
\eea
Both choices of sign have obvious physical significance for the support of the respective wave-functions, and for whether the gravitational shockwave amounts to a time delay or a time advance. The two types of energy eigenmodes are related via a unitary basis transformation
\bea
\label{basistrafo}
\bigl| \psi_+(\nu) \ra \is 
\tilde{\alpha}_{+}\raisebox{-2.5pt}{${}_{{}^{\rm R}}$}\spc  \bigl|\psi_{{}_{\rm R}}(\nu) \ra  + \tilde{\alpha}_{+}\raisebox{-2.5pt}{${}_{{}^{\rm L}}$}
\spc \bigl|\psi_{{}_{\rm L}}(- \nu) \ra,\nonumber\\[-2mm]\\[-2mm]\nonumber
\bigl| \psi_-(\nu) \ra \is 
\tilde{\alpha}_{-}\raisebox{-2.5pt}{${}_{{}^{\rm R}}$}\spc  \bigl|\psi_{{}_{\rm R}}(\nu) \ra  + \tilde{\alpha}_{-}\raisebox{-2.5pt}{${}_{{}^{\rm L}}$}
\spc \bigl|\psi_{{}_{\rm L}}(- \nu) \ra,
\eea
specified by the two-by-two unitary matrix\footnote{This computation is similar to the ones performed in \cite{Kiem:1995iy} and \cite{Hooft:2015jea,Hooft:2016itl}.}
\bea
\label{alphamatrix}
\left(\!\! \begin{array}{cc} {\tilde{\alpha}_{+}\raisebox{-2.5pt}{${}_{{}^{\rm R}}$}}\! &\! {\tilde{\alpha}_{+}\raisebox{-2.5pt}{${}_{{}^{\rm L}}$}} \\[.5mm] {\tilde{\alpha}_{-}\raisebox{-2.5pt}{${}_{{}^{\rm R}}$}} \!&\! {\tilde{\alpha}_{-}\raisebox{-2.5pt}{${}_{{}^{\rm L}}$}} \end{array}\!\! \right) \, = \, \frac  { \Gamma(\textstyle \frac 1 2 - i \nu)}{\sqrt{2\pi}}\,\,  \left(\begin{array}{cc} e^{i \frac \pi 4 + \frac \pi 2 \nu} &   e^{-i \frac {\pi}  4 -\frac \pi 2 \nu } \\[2mm]
e^{-i \frac { \pi} 4 -\frac \pi 2 \nu} & e^{i \frac \pi 4 + \frac \pi 2 \nu} \end{array}\right). 
\eea

For the study of wormhole traversability, it is most informative to consider the shockwave S-matrix elements in the left- and right energy eigenmodes \eqref{lrmodes}. Concretely, we will consider the amplitude between the two-particle initial state
\bea
\li \nu_1, \nu_3 \ra{}_{{}_{\rm AB}} \is \li \psi_{{}_{\rm A}}(\nu_1)\ra \li \phi_{{}_{\rm B}}(\nu_3)\ra, \qquad \mbox{\footnotesize A,B = L,R}. 
\eea
to the final two particle state $\li \nu_2, \nu_4 \ra{}_{{}_{\rm CD}}$ with $ \mbox{\footnotesize C,D = L,R}$. The 2-to-2 scattering matrix for given initial and final energies thus reduces to a four-by-four matrix
\be
\label{smat}
\qquad\qquad {}_{{}_{\rm CD}}\!\spc \la \nu_2, \nu_4 | \, {\cal S}\, |\nu_1, \nu_3 \ra{}_{{}_{\rm AB}}\ \qquad\qquad \mbox{\footnotesize A,B,C,D = L,R} .
\ee
We will find that all matrix elements are non-zero. 
In particular, there is a non-zero amplitude for the particles to traverse the wormhole, as indicated on the right in Figure~1. This result is an inevitable consequence of the shockwave effect in combination with Heisenberg uncertainty. If the ingoing position wave functions are supported on the Kruskal half-line, the momentum wave functions are analytic on the upper or lower complex half plane, and thus are necessarily non-zero along the whole real axis. Because the outgoing coordinate \eqref{ushift} includes a shift proportional to the ingoing momentum, the outgoing wave functions have support on both sides of the horizon.\footnote{This situation is reminiscent of the GJW protocol. One important difference, however, is that the particle traverses the black hole with probability less than one.}

To obtain the S-matrix elements \eqref{smat}, we first compute the matrix elements in the basis \eqref{pmmodes} with positive and negative Kruskal momentum, and then apply the basis transformation \eqref{basistrafo}. Since the Dray-'t Hooft S-matrix ${\cal S}$ preserves the Kruskal momentum,  in the $\pm$ basis \eqref{pmmodes}  it takes the form of a diagonal four-by-four matrix with four identical eigenvalues
\bea
\label{smatf}
{S}(\nu_2,\nu_4;\nu_1,\nu_3)   \is\!  \int^{\infty}_0\! \frac{dp_{+} } {p_{+}}  \int^{\infty}_0\! \frac{dp_{-}} {p_{-}} \,
 \, p_{+}^{i\nu_1}p_{-}^{-i\nu_3} p_{+}^{-i\nu_2} p_{-}^{i\nu_4} e^{i\kappa\, p_+p_-} \nonumber\\[-1.5mm]\\[-1.5mm]\nonumber
 \is \kappa^{i(\nu_2-\nu_1)} e^{\frac \pi 2  (\nu_2-\nu_1)}\, {\Gamma\bigl(i(\nu_1-\nu_2)\bigr)}\,\spc  2\pi\spc \delta(\nu_1\! - \!\spc \nu_2\! +\! \spc \nu_3\! -\! \spc \nu_4), 
\eea
where $\kappa \equiv \frac{\beta}{4\pi C}$. 
This reduced S-matrix satisfies the unitarity relation
\bea
\int\! \frac{d\nu_2}{2\pi}\int \!\spc \frac{d\nu_4}{2\pi}\, {S}^\dagger(\nu_5,\nu_6;\nu_2,\nu_4)\spc {S}(\nu_2,\nu_4;\nu_1,\nu_3)\! \is\! (2\pi)^2 \delta(\nu_1\! -\! \spc \nu_5)\delta(\nu_3\! -\! \spc\nu_6).
\eea 
The full unitary four-by-four S-matrix \eqref{smat} in the left- and right mode basis \eqref{lrmodes} thus takes the form 
\bea
\label{sfirst}
{}_{{}_{\rm CD}}\!\spc \la \nu_2, \nu_4 | \, {\cal S}\, |\nu_1, \nu_3 \ra{}_{{}_{\rm AB}}  
\is \sum_{s,s'}   \; \tilde{\alpha}_{\mbox{\tiny C} s}(\nu_2) \, \tilde{\alpha}^{\dag}{\!\!}_{s \mbox{\tiny  A} }(\nu_1)   \,  \tilde{\alpha}_{\mbox{\tiny D}s'}(\nu_4)\, \tilde{\alpha}^{\dag}{\!\!}_{s' \mbox{\tiny  B} }(\nu_3)  \; {S}(\nu_2,\nu_4;\nu_1,\nu_3) \;   .
\eea
The explicit form of this S-matrix is found by inserting the explicit expressions for ${S}(\nu_2,\nu_4;\nu_1,\nu_3)$ given in \eqref{smatf} and for $\tilde{\alpha}_{s  \mbox{\tiny  A}}(\nu_i)$ given in \eqref{alphamatrix}. We will not write it out here.

\begin{figure}[t]
\begin{center}
\includegraphics[width=0.7\textwidth]{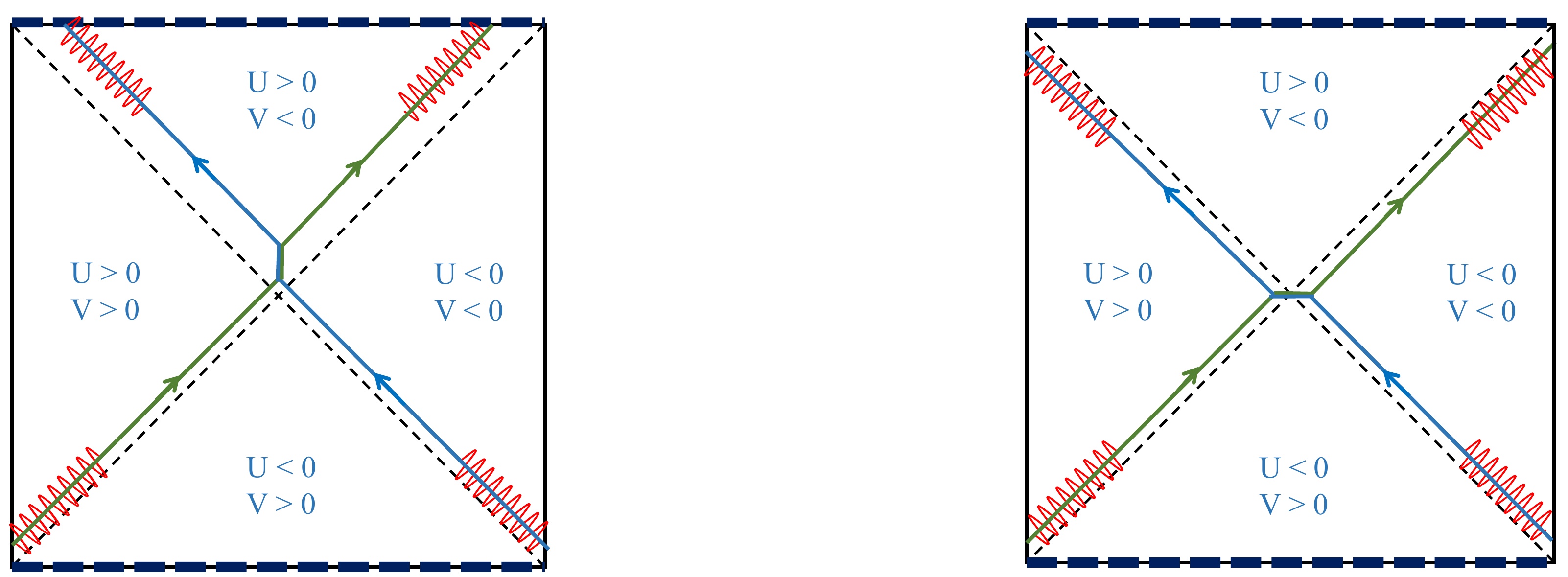}
\end{center}
\caption{The 2-to-2 shockwave scattering matrix has non-zero matrix elements between incoming and outgoing pairs of waves on either side of the black hole horizon. The interaction can produce a time delay (left) or a time advance (right). 
In the latter case, the particles traverse the wormhole. 
}
\label{shocks}
\end{figure}
\vspace{-1mm}

\subsection{Second quantized S-matrix}
\label{sect:secondq}

\vspace{-2mm}

Indeed, this is not yet the end of the story. Our discussion thus far has been within a first quantized setting. We would like to translate the above first quantized 2-to-2 particle S-matrix into a statement about the gravitational shockwave interaction between modes of an (otherwise freely propagating) quantum field.

Within second quantization, we need the relevant Bogoliubov transformations relating modes with definite localization in either the $L$- or $R$-wedge to modes containing only positive (or negative) Kruskal momentum. This is the standard construction of Unruh modes as in e.g. \cite{Unruh:1976db}. We collect the relevant formulas in appendix \ref{app:Unruh}.
We will denote the corresponding creation and annihilation operators by $a^\dagger_{{}_{{}^{\rm L,R}}}$ and $a_{{}_{{}^{\rm L,R}}}$. The Kruskal vacuum state is given by the thermo-field double state in the Rindler basis, and is annihilated by the annihilation operators $c^1(\nu)$ and $c^2(\nu)$ associated with the modes with positive Kruskal energy: 
\bea
\label{vacco}
c^1(\nu)\left|\text{TFD}\right\rangle = c^2(\nu)\left|\text{TFD}\right\rangle = 0.
\eea
A quantum field mode, when acting on the thermo-field double state, cannot carry any negative Kruskal momentum. This has obvious significance for our discussion, as it appears to eliminate the possibility that the shockwave due to a single particle can be sourced by negative Kruskal momentum. However, as we will see, the amplitude for traversing the wormhole remains non-zero.

Using the vacuum condition \eqref{vacco} and the Bogoliubov relations \eqref{bogoliubov}
\bea
a^\dagger_{{}_{{}^{\rm R}}}(\nu)  \is \alpha_{+}\raisebox{-2.5pt}{${}_{{}^{\rm R}}$}(\nu) \, c^{1\dagger}(\nu)
+  \alpha_{-}\raisebox{-2.5pt}{${}_{{}^{\rm R}}$}(\nu) \, c^2(\nu) ,
\eea
we can express the action of the Rindler creation operators $a^\dagger_{{}_{{}^{\rm R}}}$ 
on the right wedge in terms of the action of a creation operator $c^{1\dagger}$ with positive Kruskal momentum via
\bea
a^\dagger_{{}_{{}^{\rm R}}}(\nu) \spc \li\mbox{\small TFD}\ra \is \alpha_{+}\raisebox{-2.5pt}{${}_{{}^{\rm R}}$}(\nu) \, c^{1\dagger}(\nu)
 \li \mbox{\small TFD}\ra .
 \eea
Analogous equations hold for the other creation and annihilation operators.

In the second quantized setting,
the scattering amplitude on the right of Figure 1 is given by
\bea
 {}_{{}_{\rm LR}}\!\spc 
\la \nu_2, \nu_4 | \, {\cal S}\, |\nu_1, \nu_3 \ra{}_{{}_{\rm RL}}
\is 
 \la \mbox{\small TFD} \ri\spc a_{{}_{{}^{\rm L}}}\! (\nu_2) \, a_{{}_{{}^{\rm R}}}\!(\nu_4) \; {\cal S}  \; a_{{}_{{}^{\rm R}}}^\dag\!(\nu_1)\spc a_{{}_{{}^{\rm L}}}^\dag\!(\nu_3)\li \mbox{\small TFD} \ra .
   \eea 
From the above discussion, it is clear that this matrix element is {\it not} the same as the matrix element computed in the first quantized theory. Instead, it is given by keeping only the subcomponent ${\cal S}_{++}$ of the unitary 2-to-2 scattering matrix ${\cal S}$ that acts within the positive Kruskal momentum sector. In other words, the matrix element is found by keeping only the $s,s' = ++$ 
   term in the expression \eqref{sfirst} for the first quantized S-matrix.\footnote{We of course also replace the unitary matrix $\tilde{\alpha}_{\pm}\raisebox{-2.5pt}{${}_{{}^{\rm R,L}}$}$ with the Bogoliubov matrix $\alpha_{\pm}\raisebox{-2.5pt}{${}_{{}^{\rm R,L}}$}$.} Inserting the explicit expressions \eqref{smatf} and \eqref{bogoliubov} for ${S}(\nu_2,\nu_4;\nu_1,\nu_3)$ and $\alpha_{s  \mbox{\tiny  A}}(\nu_i)$ yields our final answer for the scattering amplitude in which the two particles traverse the wormhole
 \bea {}_{{}_{\rm LR}}\!\spc \la \nu_2, \nu_4 \ri  \, {\cal S}\, \li \nu_1, \nu_3 \ra{}_{{}_{\rm RL}}
  \is  \hspace{-.5mm}\frac 1{4\pi^2} \,  e^{\frac \pi 2 (\nu_1\nspc - \nspc \nu_2\nspc + \nu_3\nspc - \nu_4)}\, \textstyle {\Gamma(1 +\nspc i\nu_1)\Gamma(1 -\nspc i\nu_2)\Gamma(1 +\nspc i\nu_3)\Gamma(1 -\nspc i\nu_4)}\qquad\\[2.5mm]
 && \  \times \ \kappa^{i(\nu_2-\nu_1)} \, e^{\frac \pi 2 (\nu_2 - \nu_1)} \, \Gamma\(i(\nu_1-\nu_2)\) \; 2\pi  \delta(\nu_1\! -\!\spc \nu_2\! +\! \spc \nu_3\! -\! \spc \nu_4).\qquad\qquad
  \nonumber 
\eea 

Does this non-zero result imply that shockwave interactions enable information to travel faster than the speed of light? The cautious answer is: no. First, note that even without interactions, the wormhole traversing amplitude is non-zero ${}_{{}_{\rm LR}}\!\spc \langle \nu_2, \nu_4  |\nu_1, \nu_3 \rangle{}_{{}_{\rm RL}} = \frac{\pi \nu_1}{{\rm sinh} \pi \nu_1 } \frac{\pi \nu_3}{{\rm sinh} \pi \nu_3 } \; \delta(\nu_{12}) \delta(\nu_{34})$. Still we know that free field theory is perfectly causal, and space-like separated operators commute. Secondly, we observe that the ratio between the wormhole traversing and non-traversing amplitude is given by a simple thermal factor
\bea
\frac{ {}_{{}_{\rm LR}}\!\spc \la \nu_2, \nu_4 | \, {\cal S}\, |\nu_1, \nu_3 \ra{}_{{}_{\rm RL}}}{ {}_{{}_{\rm RL}}\!\spc \la \nu_2, \nu_4 | \, {\cal S} |\nu_1, \nu_3 \ra{}_{{}_{\rm RL}}} \is e^{-{\pi}(\nu_2+\nu_4)} .
\eea
This means that in position space, the wormhole traversing amplitude\footnote{These integrals in \eqref{positionamp} can be explicitly evaluated in terms of the confluent hypergeometric function \cite{Maldacena:2016upp}
\bea
\frac{\la V_1^{{}_{{}^{\rm L}}}W_3^{{}_{{}^{\rm R}}}V_2^{{}_{{}^{\rm R}}}W_4^{{}_{{}^{\rm L}}}\ra}{\langle V_1^{{}^{{}_{\rm L}}} V_2^{{}^{{}_{\rm R}}}\rangle\langle W_3^{{}^{{}_{\rm R}}} W{}^{{}^{{}_{\rm L}}}_4\rangle} \!\!  \is \!\! \frac 1 z \; U(1, 1 , 1/z) \qquad \qquad z \, = \,\frac{i\, e^{( t_{3}+t_4-t_{1}-t_2)/2}} {4\kappa \, \cosh \frac{t_{12}}{2} \cosh\frac {t_{34}} 2}
\eea
with $U(1,1,x) = \int_0^\infty \!\! ds \, \frac{e^{-xs}}{1+s}$ and $z$ the cross ratio. This amplitude is a smooth function of all time coordinates.}
\bea
\label{positionamp}
\la V_1^{{}_{{}^{\rm L}}} W_3^{{}_{{}^{\rm R}}}  V_2^{{}_{{}^{\rm R}}}  W_4^{{}_{{}^{\rm L}}} \ra \, = \, \prod_{i=1}^4\int_{-\infty}^\infty\!\!\spc \frac{d\nu_i}{2\pi}\; e^{i(\nu_1 t_1-\nu_2 t_2+\nu_3 t_3-\nu_4 t_4)}  {}_{{}_{\rm LR}}\!\spc \la \nu_2, \nu_4 | \, {\cal S}\, |\nu_1, \nu_3 \ra{}_{{}_{\rm RL}}
\eea
is obtained from the non-traversing amplitude $
\langle V_1^{{}_{{}^{\rm L}}} W_3^{{}_{{}^{\rm R}}}  V_2^{{}_{{}^{\rm L}}}  W_4^{{}_{{}^{\rm R}}} \rangle$ via a simple analytic continuation $(t_2,t_4) \to (t_2 + i \pi ,t_4 + i \pi)$  by an imaginary shift  in the time coordinates of the outgoing particles. The physical interpretation of the above wormhole traversing 2-to-2 amplitude is that it is simply the imprint of the {\it causal} shockwave interaction onto the non-local EPR correlations in the thermo-field double state. Hence it does not correspond to acausal signal propagation and does not by itself give rise to non-zero commutators between space-like separated local operators.  The implementation of a GJW type protocol indeed requires a more elaborate set up than we have considered here.

\section{Schwarzian correlation functions}

\label{sect3}

\vspace{-2mm}

In this section we give a brief summary of the exact solution of the Schwarzian quantum mechanics, and present the explicit expression of the two- and four-point functions. A more detailed discussion can be found in \cite{MTV} and \cite{thomas}. In the next two sections, we will then take the large $C$ limit of these results and establish a precise match with the shockwave scattering. 

\vspace{-1mm}

\subsection{Schwarzian QM as a limit of Virasoro CFT}

\vspace{-2mm}

Schwarzian quantum mechanics at finite temperature is defined as a functional integral over the coset Diff$(S^1)/SL(2,\mathbb{R})$ of the group of diffeomorphisms Diff$(S^1)$
of the thermal circle
\bea
f: \quad \tau & \to &  f(\tau) ,\qquad  \qquad  f(\tau+2\pi) = f(\tau) + 2\pi,
\eea 
modulo the group $SL(2,\mathbb{R})$ of M\"obius transformations 
\bea
F \to \frac{a F + b}{c F + d},
\eea
acting on the periodic function $F = \tan f/2$. For this subsection only, we choose units such that the inverse temperature $\beta=2\pi$. 

This geometric fact, that the space of all paths in the functional integral forms a coset of two familiar groups, enables one to use some powerful machinery for solving its correlation functions \cite{StWi, MTV, thomas}.
Diff$(S^1)$ is also known as the Virasoro group and $\frac{\rm Diff(S^1)}{SL(2,\mathbb{R})}$ is called the special coadjoint orbit of the identity element. A coadjoint orbit of any continuous group admits a natural symplectic structure, that via the standard quantization rules gives rise to commutation relations among the Noether charges that equal the Lie algebra of the infinitesimal symmetry transformations.  For $\frac{\rm Diff(S^1)}{SL(2,\mathbb{R})}$  this algebra takes the form of the Virasoro algebra with central charge $c$ related to the quantization parameter $\hbar$ via $\hbar = \frac{24\pi}{c}$. The Hilbert space of the corresponding quantum theory is given by the identity module of the Virasoro algebra.

Suppose we know how to perform a field redefinition from $f(\tau)$ to some new set of canonically conjugate variables $\phi(\tau)$ and $\pi_\phi(\tau)$, such that, upon introducing $\hbar$, it implies the commutation relation $[\phi(\tau_1), \pi_\phi(\tau_2)] = i\hbar \delta(\tau_{12})$. Consider now the Lagrangian
\bea
\label{ltd}
L _{2D} \is \int_{-\pi}^\pi\!\!\! d\tau\, \bigl(\pi_\phi \dot \phi + \{ F, \tau\} \bigr), 
\eea
where $\dot \phi = \partial_s \phi$ denotes the derivative with respect to an auxiliary extra time variable $s$, and 
where $\pi_\phi, \phi$ and $F = \tan \frac f 2$ all denote periodic functions on the interval $-\pi < \tau < \pi$. By definition, the quantization of this theory produces a Hilbert space given by the identity module ${\cal H}_0$ of the Virasoro algebra with central charge $c = \frac{24\pi}{\hbar}$.

Let us now place the 2D theory (\ref{ltd}) on a small periodic time interval $0<s<T$ with period $T$. Performing the functional integral computes the partition function $Z = \tr_{{\cal H}_0}(e^{-T H}\bigr)$ with
\bea
\label{hamil}
H \is  
-  \int_{-\pi}^\pi \!\! d\tau\, \left\{F(\tau,s),\tau\right\} , \qquad \quad F = \tan \frac f 2 .
\eea
The exact computations of \cite{MTV} are based on the observation that the Schwarzian theory is obtained from the above 2D theory by taking the combined $\hbar \to 0$ and $T\to 0$ limit with  $T/\hbar = \frac{cT}{24\pi}=C$  held fixed. The theory then dynamically reduces to just the zero-mode along the $s$-direction
\bea
\label{phasint}
\hspace{-0.5cm}\int\displaylimits_{\tiny 
f(\pi,s) = f(-\pi,s)+2\pi} \hspace{-2.5em} \mathcal{D}\nspc f \; e^{\frac{1}{\hbar}\int\limits_{-\pi}^{\pi}\! d\tau \int_{0}^{T} ds \left(i\pi_\phi \dot{\phi} +\left\{F,\tau\right\} \right)}
\hspace{.5em}
& \to & \hspace{.5em} \int\displaylimits_{\tiny
f(\pi) = f(-\pi)+2\pi} \hspace{-2.6em} \mathcal{D}\nspc f\; e^{\, C\! \int\limits_{-\pi}^{\pi} d\tau\, \left\{F,\tau\right\} }. 
\eea
This identity holds under the condition that the functional integration measure on both sides is defined in terms of the symplectic form on Diff$(S^1)$. 

The 2D theory introduced above turns out to be equivalent to Liouville CFT defined on the strip $0< \tau < \pi$ with ZZ-brane boundary conditions \cite{Zamolodchikov:2001ah} at $\tau =0$ and $\tau=\pi$. Here we briefly ouline the argument \cite{thomas}.
Liouville  CFT with a boundary is defined by the Hamiltonian 
\bea
\label{hliou}
{H} =  \int_0^\pi \!\! d\tau \, \Bigl( \, \frac{ \pi_\phi^2}{2} \spc + \spc \frac{\phi_\tau^2}{2}\spc + \spc e^{\phi} -\spc 2\phi_{\tau\tau}\Bigr),
\eea
where $\pi_\phi$ and $\phi$ are canonically conjugate fields. To establish this fact, we perform the following field redefinition, first introduced by Gervais and Neveu \cite{Gervais:1981gs,Gervais:1982nw,Gervais:1982yf,Gervais:1983am} 
\bea
\label{gntrafo}
e^{\phi} = -8 \frac{A_\tau B_\tau}{(A-B)^2}    \ \ & ; & \ \ 
\pi_\phi = \frac{A_{\tau\tau}}{A_\tau} - \frac{B_{\tau\tau}}{B_\tau}-2\frac{A_\tau+B_\tau}{(A-B)},
\eea
where $A_\tau = \partial_\tau A$ etc. In terms of these new variables, the ZZ-boundary conditions impose that $A(0) = B(0)$ and $A(\pi) = B(\pi)$. This allows the introduction of a doubled field variable $f(\tau)$ 
\bea
\label{abf}
            A(\tau)\, = \, \spc \tan    \frac {f(\tau)}2\spc , \qquad & &\ 0<\tau<\pi, \nonumber \\[-2mm]\\[-2mm]\nonumber
              B(-\tau)  = \, \tan \frac {f(\tau)}2, \qquad & & -\pi < \tau < 0,
\eea
which defines a continuous function with periodic boundary conditions $f(\tau + 2\pi) = f(\tau)+2\pi$.  The above two equations \eqref{gntrafo} and \eqref{abf} specify the mapping from the canonical $(\phi,\pi_\phi)$ variables to the Diff$(S^1)$ variable $f(\tau)$.
It is easy to verify that the Liouville Hamiltonian \eqref{hliou}, when expressed in terms of $f(\tau)$, takes the form (\ref{hamil}). Moreover -- and this is an essential element in the construction -- the canonical Poisson bracket relation between $\phi$ and $\pi_\phi$ transforms via the above field redefinition precisely into the correct Virasoro symplectic form on Diff$(S^1)$. This confirms that the two 2D theories are indeed the same.

The classical solution of Liouville field with the above ZZ-brane boundary conditions is expressed 
in terms of the single function $f$  via \cite{Dorn:2006ys,Dorn:2008sw}
\bea
\label{phisol}
e^{\phi(u,v)} \is \frac{2 f'(u) f'(v)}{\sin\left(\frac{f(u)-f(v)}{2}\right)^2},
\eea
where $u= s+t$ and $v=s-t$ denote the light-cone coordinates. This formula for the classical field will be useful in the next subsection.

\vspace{-1mm}
\subsection{Feynman rules in the Schwarzian theory}
\label{sectFR}

\vspace{-2mm}
In the following sections, we will study $2n$-point functions in Schwarzian QM of the form
\bea
\Bigl\langle \, \prod_{i=1}^n V_i(\tau_{2i})V_i(\tau_{2i+1}) 
 \, \Bigr\rangle \is \int
\! {{\cal D}\!\spc f}\, e^{- S[f] }\,  \prod_{i=1}^n  {\cal O}_{\ell_i}(\tau_{2i},\tau_{2i+1}),  
\eea
where ${\cal O}_\ell(\tau_1,\tau_2)$ denote the $SL(2,\mathbb{R})$ invariant bi-local operators 
\bea
{\cal O}_\ell(\tau_1,\tau_2) \is \left( \frac{\sqrt{2 f'(\tau_1)f'(\tau_2)}}{\sin \frac{1}{2}[f(\tau_1)-f(\tau_2)]} \right)^{2\Deltal}. 
\eea
In the microscopic SYK theory, each bi-local operator ${\cal O}_\ell(\tau_1,\tau_2)$ represents the insertion of a pair $V_\ell(\tau_1) \, V_\ell(\tau_2)$ of local scaling operators with scale dimension $\ell$. 

In the dictionary between Schwarzian QM and Virasoro CFT, ${\cal O}_\ell(\tau_1,\tau_2)$ gets mapped to the exponential vertex operators $V_\ell(\tau_1,\tau_2) = e^{\ell \phi(\tau_1,\tau_2)}$. The  derivation of this map makes use of the classical solution \eqref{phisol} of the Liouville field in the presence of the two ZZ-branes. In \cite{MTV}, exact expressions for the Schwarzian $2n$-functions were obtained by taking a suitable limit of known results from Virasoro CFT. In this subsection, we will just quote the results. For the derivations and some checks, we refer to \cite{MTV}. In the next three sections, we will then compare the exact formulas with the semi-classical bulk expectations.

General finite temperature correlation functions of $n$ bi-local operators in Schwarzian quantum mechanics can be written in the form of a multi-dimensional `momentum integral'
\bea
\Bigl\langle \, \prod_{i=1}^n V_i(\tau_{2i})V_i(\tau_{2i+1}) 
 \, \Bigr\rangle \is \ \prod_{i}\int d k_i^2\sinh 2\pi k_i\  {\cal A}_{2n}(k_i,\tau_i),
\eea
where $d k^2 = 2k dk$. Here each $k_i$ labels a complete set of intermediate energy eigenstates weighted by the appropriate spectral density 
\bea
\mathbf{1} = \int\! dk^2 \,  \sinh 2\pi k   \; | k \rangle \langle k |, \qquad \qquad E = \frac{k^2}{2C}.
\eea
The momentum space amplitudes for each $2n$-point function can immediately be written down by applying the following simple set of Feynman rules.  

Every graph is circumscribed by a circle, which represents the thermal circle. Inside the circle, we draw a line for every bi-local operator, which connects the corresponding two points on the boundary circle. Propagators on the circular boundary and vertices are given in terms of Euclidean time by
\bea 
\label{frules}
\begin{tikzpicture}[scale=0.57, baseline={([yshift=-0.1cm]current bounding box.center)}]
\draw[thick] (-0.2,0) arc (170:10:1.53);
\draw[fill,black] (-0.2,0.0375) circle (0.1);
\draw[fill,black] (2.8,0.0375) circle (0.1);
\draw (3.4, 0) node {\footnotesize $\tau_1$};
\draw (-0.7,0) node {\footnotesize $\tau_2$};
\draw (1.25, 1.6) node {\footnotesize $k$};
\draw (6.5, 0) node {$\raisebox{6mm}{$\ \ =\ \ e^{-\spc \spc \frac {k^2}{2C}  \spc (\tau_2-\tau_1)}$}$};
\end{tikzpicture}, ~~~~~~~~~~\ \ \begin{tikzpicture}[scale=0.7, baseline={([yshift=-0.1cm]current bounding box.center)}]
\draw[thick] (-.2,.9) arc (25:-25:2.2);
\draw[fill,black] (0,0) circle (0.08);
 \draw[thick](-1.5,0) -- (0,0);
\draw (.3,-0.95) node {\footnotesize $\textcolor{black}{k_2}$};
\draw (.3,0.95) node {\footnotesize$\textcolor{black}{k_1}$};
\draw (-1,.3) node {\footnotesize$\Deltal$};
\draw (2.5,0.1) node {$\mbox{$\ =\  \, \gamma_\ell(k_1,k_2)\spc .$}$}; \end{tikzpicture}\
\eea
 The time dependent  factor represents the usual Schr\"odinger time evolution of the intermediate energy eigenstates. The vertex factor takes the following form
\beq
\label{gammaell}
\gamma_\ell(k_1,k_2)\; = \; \sqrt{\frac{ \Gamma(\Deltal\pm i k_1\pm i k_2)\!}{\Gamma(2\Deltal)}}.
\eeq
This vertex factor represents the matrix element of each endpoint of the bi-local operator between the corresponding two energy eigenstates.

The lines in the diagram divide the disk into several disjoint sectors. In order to write down a general amplitude, one associates a separate momentum label to each region of the disk, even those that do not reach the boundary circle. Then for each crossing of lines within the diagram, one writes the relevant $R$-matrix or crossing matrix, labeled by the four momenta of the four regions defined by these crossing lines and the two momenta attached to the lines. In diagrammatic notation, we write
\bea
\label{crossing}
\raisebox{15pt}{ \begin{tikzpicture}[scale=.6, baseline={([yshift=0cm]current bounding box.center)}]
\draw[thick] (-0.85,0.85) -- (-.15,.15);
\draw[thick] (.15,-.15) -- (0.85,-0.85);
\draw[thick] (-0.85,-0.85) -- (0.85,0.85);
\draw[dotted,thick] (-0.85,-0.85) -- (-1.25,-1.25);
\draw[dotted,thick] (0.85,0.85) -- (1.25,1.25);
\draw[dotted,thick] (-0.85,0.85) -- (-1.25,1.25);
\draw[dotted,thick] (0.85,-0.85) -- (1.25,-1.25);
\draw (1.5,0) node {\scriptsize $k_s$};
\draw (-1.5,0) node {\scriptsize $k_t$};
\draw (-.75,.33) node {\scriptsize $\ell_2$};
\draw (.78,.33) node {\scriptsize $\ell_1$};
\draw (0,1.5) node {\scriptsize  $k_1$};
\draw (0,-1.5) node {\scriptsize $k_4$};
\end{tikzpicture}~~\raisebox{-3pt}{$\ \ \  = \ \ R_{k_sk_t}\! 
 \left[\, {}^{k_4}_{k_1} \,{}^{\Deltal_2}_{\Deltal_1}\right] $}~~~}
\eea
The R-matrix $ R_{k_sk_t}\! 
 \left[\, {}^{k_4}_{k_1} \,{}^{\Deltal_2}_{\Deltal_1}\right]$ is equal to the $6j$-symbol of $SU(1,1)$, whose explicit form derived in \cite{MTV} is given in equation \eqref{appfinal}. 

It is perhaps instructive to elaborate on how this Feynman rule prescription is deduced from the Virasoro CFT.  
In the dictionary of \cite{MTV}, each vertex operator in 2D Virasoro CFT maps to a bilocal  operator in the 1D Schwarzian theory. As a concrete example, let us consider the three-point function in 2D, or the six-point function in 1D
\begin{equation}
\label{6ptdiagr}
\Bigl\langle V_{\ell_1}(z_1,\bar{z}_1)V_{\ell_2}(z_2,\bar{z}_2)V_{\ell_3}(z_3,\bar{z}_3) \Bigr\rangle \ \ = \ \begin{tikzpicture}[scale=0.65, baseline={([yshift=0cm]current bounding box.center)}]
\draw[thick] (0,0) circle (1.5);
\draw[thick] (1.06,1.06) arc (300:240:2.16);
\draw[thick] (-1.06,-1.06) arc (120:60:2.16);
\draw[fill,black] (-1.06,-1.06) circle (0.1);
\draw[thick] (-1.5,0)--(1.5,0);
\draw (-1.5,-1.5) node {\small $\tau_4$};
\draw (-1.5,1.5) node {\small $\tau_2$};
\draw (1.5,-1.5) node {\small $\tau_5$};
\draw (1.5,1.5) node {\small $\tau_1$};
\draw (2.2,0) node {\small $\tau_6$};
\draw (-2.2,0) node {\small $\tau_3$};
\draw[fill,black] (1.06,-1.06) circle (0.1);
\draw[fill,black] (-1.06,1.06) circle (0.1);
\draw[fill,black] (1.06,1.06) circle (0.1);
\draw[fill,black] (1.5,0) circle (0.1);
\draw[fill,black] (-1.5,0) circle (0.1);
\draw (0,1.1) node {\small $\ell_1$};
\draw (0,.3) node {\small $\ell_2$};
\draw (0,-1.1) node {\small $\ell_3$};
\end{tikzpicture}
\end{equation}
where the time arguments of the bilocal operators are identified with the 2D locations via: \ \ 
$(z_1,\bar{z}_1) \to (\tau_2, \tau_1), \quad (z_2,\bar{z}_2) \to (\tau_3, \tau_6), \quad (z_3,\bar{z}_3) \to (\tau_4, \tau_5) $.

The Schwarzian limit amounts to taking the length of the cylinder between operator insertions to be infinitely long. In this limit, only primaries propagate in the intermediate channels. Inserting a complete sets of states thus amounts to the leading OPE expansion
\begin{equation}
\label{ope}
\left\langle O_i(\infty) V_{\ell} (z) O_j(0)\right\rangle = z^{h_i - h_{\ell} - h_j} c_{i \ell j}.
\end{equation}
The large $c$ limit of the DOZZ OPE coefficients $c_{i\ell j}$ reduces to  the vertex functions $\gamma_{\ell}(k_i,k_j)$, while the $z$-dependent pieces in \eqref{ope} become propagators along the cylinder between insertions.

To obtain the OTO six-point function, we start from the above striped diagram and swap first $\tau_2$ with $\tau_3$ and then $\tau_2$ with $\tau_4$. Swapping operators is achieved by using the Schwarzian limit of the braiding $R$-matrix of Virasoro CFT, which was determined by Ponsot-Teschner in \cite{PT}. So we only need to swap the first argument $\tau_2$ all the way through all of the first arguments of the remaining Liouville operators. This means that this acts fully in the holomorphic sector. As Liouville vertex operators naturally factorize in chiral and anti-chiral parts, we can just use the exchange algebra of chiral operators, 
$V_{\ell_1}(z_1) V_{\ell_2}(z_2) = R_{12} V_{\ell_2}(z_2) V_{\ell_1}(z_1)$, 
to move the holomorphic part of $V_{\ell_1}$ all the way to the other side.

The double shockwave process can now be found by doing a two-move process
\begin{align}
\begin{tikzpicture}[scale=0.63, baseline={([yshift=0cm]current bounding box.center)}]
\draw[thick] (0,0) circle (1.5);
\draw[thick] (1.06,1.06) arc (300:240:2.16);
\draw[thick] (-1.06,-1.06) arc (120:60:2.16);
\draw[thick] (-1.5,0)--(1.5,0);
\draw (-1.5,-1.5) node {\small $\tau_4$};
\draw (-1.5,1.5) node {\small $\tau_2$};
\draw (1.5,-1.5) node {\small $\tau_5$};
\draw (1.5,1.5) node {\small $\tau_1$};
\draw (2.2,0) node {\small $\tau_6$};
\draw (-2.2,0) node {\small $\tau_3$};
\draw[fill,black] (-1.06,-1.06) circle (0.1);
\draw[fill,black] (1.06,-1.06) circle (0.1);
\draw[fill,black] (-1.06,1.06) circle (0.1);
\draw[fill,black] (1.06,1.06) circle (0.1);
\draw[fill,black] (1.5,0) circle (0.1);
\draw[fill,black] (-1.5,0) circle (0.1);
\draw (0,1.1) node {\small $\ell_1$};
\draw (0,.3) node {\small $\ell_2$};
\draw (0,-1.1) node {\small $\ell_3$};
\draw (0,1.9) node {\small \color{red}$k_1$};
\draw (0,-1.9) node {\small \color{red}$k_2$};
\draw (1.85,-.7) node {\small \color{red}$k_j$};
\draw (1.85,.7) node {\small \color{red}$k_i$};
\draw (-1.85,.7) node {\small \color{red}$k_i$};
\draw (-1.85,-.7) node {\small \color{red}$k_j$};
\end{tikzpicture}~~\raisebox{-3pt}{$\  \to \ \ R_{k_ik_{k}}\! 
 \left[\, {}^{k_j}_{k_1} \,{}^{\ell_2}_{\ell_1}\right] $}~~~
\begin{tikzpicture}[scale=0.6, baseline={([yshift=0cm]current bounding box.center)}]
\draw[thick]  (0,0) ellipse (1.6 and 1.6);
\draw[thick] (1.05,1.2) arc (-35.7955:-93.2014:3);
\draw[thick] (-1.05,1.2) arc (-144.2045:-86.7986:3);
\draw[thick] (-1.13,-1.13) arc (120:60:2.19);
\draw (-1.5,-1.5) node {\small $\tau_4$};
\draw (-1.5,1.5) node {\small $\tau_3$};
\draw (1.5,-1.5) node {\small $\tau_5$};
\draw (1.5,1.5) node {\small $\tau_1$};
\draw (2.2,0) node {\small $\tau_6$};
\draw (-2.2,0) node {\small $\tau_2$};
\draw[fill,black] (-1.6,0) circle (0.1);
\draw[fill,black] (-1.05,1.2) circle (0.1);
\draw[fill,black] (1.6,0) circle (0.1);
\draw[fill,black] (1.05,1.2) circle (0.1);
\draw[fill,black] (-1.13,-1.13) circle (0.1);
\draw[fill,black] (1.13,-1.13) circle (0.1);
\draw (0.45,1.15) node {\small $\ell_1$};
\draw (-0.45,1.15) node {\small $\ell_2$};
\draw (-0.1,-.6) node {\small $\ell_3$};
\draw (0,2) node {\small \color{red}$k_1$};
\draw (0,-2) node {\small \color{red}$k_2$};
\draw (1.97,-.75) node {\small \color{red}$k_j$};
\draw (1.97,.75) node {\small \color{red}$k_i$};
\draw (-1.97,.75) node {\small \color{red}$k_{k}$};
\draw (-1.97,-.75) node {\small \color{red}$k_{j}$};
\end{tikzpicture} \nonumber \\
~~\raisebox{-3pt}{$\  \to \ \ R_{k_ik_{k}}\! 
 \left[\, {}^{k_j}_{k_1} \,{}^{\ell_2}_{\ell_1}\right] R_{k_jk_{l}}\! 
 \left[\, {}^{k_2}_{k_{k}} \,{}^{\ell_3}_{\ell_1}\right]$}~~~
\begin{tikzpicture}[scale=0.6, baseline={([yshift=0cm]current bounding box.center)}]
\draw[thick]  (0,0) ellipse (1.6 and 1.6);
\draw[thick]  plot[smooth, tension=.7] coordinates {(1.6,0) (-1.05,1.2)};
\draw[thick]  plot[smooth, tension=.7] coordinates {(1.13,-1.13) (-1.6,0)};
\draw[thick]  plot[smooth, tension=.7] coordinates {(1.05,1.2) (-1.13,-1.13)};
\draw (-1.5,-1.5) node {\small $\tau_2$};
\draw (-1.5,1.5) node {\small $\tau_3$};
\draw (1.5,-1.5) node {\small $\tau_5$};
\draw (1.5,1.5) node {\small $\tau_1$};
\draw (2.2,0) node {\small $\tau_6$};
\draw (-2.2,0) node {\small $\tau_4$};
\draw[fill,black] (-1.6,0) circle (0.1);
\draw[fill,black] (-1.05,1.2) circle (0.1);
\draw[fill,black] (1.6,0) circle (0.1);
\draw[fill,black] (1.05,1.2) circle (0.1);
\draw[fill,black] (-1.13,-1.13) circle (0.1);
\draw[fill,black] (1.13,-1.13) circle (0.1);
\draw (0.45,1.15) node {\small $\ell_1$};
\draw (-0.75,0.7) node {\small $\ell_2$};
\draw (0.9,-.6) node {\small $\ell_3$};
\draw (0,2) node {\small \color{red}$k_1$};
\draw (0,-2) node {\small \color{red}$k_2$};
\draw (1.97,-.75) node {\small \color{red}$k_j$};
\draw (1.97,.75) node {\small \color{red}$k_i$};
\draw (-1.97,.75) node {\small \color{red}$k_{k}$};
\draw (-1.97,-.75) node {\small \color{red}$k_{l}$};
\end{tikzpicture}
\end{align}

This can be generalized to higher-point functions in a tedious but straightforward way. We comment on it in appendix \ref{secthigh}.  

\section{Two-point function} 
\label{sect4}
\vspace{-2.5mm}

In this section we will study the large $C$ limit of the Lorentzian two-point function of the Schwarzian theory. 
The exact answer for the two point function found in \cite{MTV} reads 
\bea\label{2ptExact}
\la V_1(0) V_2(t) \ra \is \prod_{i=1,2} \int \! dk^2_i \spc \sinh 2\pi k_i \; \;
\raisebox{3pt}{\begin{tikzpicture}[scale=0.55, baseline={([yshift=0cm]current bounding box.center)}]
\draw[thick] (0,0) circle (1.5);
\draw[thick] (-1.5,0) -- (1.5,0);
\draw[fill,black] (-1.5,0) circle (0.1);
\draw (0.05,1.15) node {\footnotesize $k_1$};
\draw[fill,black] (1.55,0) circle (0.1);
\draw (0.05,-1.15) node {\footnotesize $k_2$};
\draw (-0,.3) node {\footnotesize $\ell$};
\end{tikzpicture}}\\[2mm]
\is \prod_{i=1,2} \int \! dk^2_i \spc \sinh 2\pi k_i \; \; e^{-\frac{i}{2C}(k_1^2 - k^2_2) t- \frac{\beta}{2C} k_2^2}\  \frac{\Gamma(\ell \pm i(k_1 \pm k_2))}{(2C)^{2\ell}\, \Gamma(2\ell)},
\eea
where the $\pm$ notation denotes the products of all choices of signs\footnote{For example $\Gamma(a\pm b\pm c) = \Gamma(a+b+c) \Gamma(a+b-c) \Gamma(a-b+c) \Gamma(a-b-c)$.}. The time dependent phase factor represents the usual Schr\"odinger evolution of the intermediate energy eigenstates. 

As a preparation for the discussion of the OTO four point function, we introduce the asymptotic wave functions that describe the scattering states in AdS${}_2$. We extract these wavefunctions from the boundary-to-boundary propagator and from the large $C$ limit of the two-point function.

\vspace{-1mm}
\subsection{Asymptotic wavefunctions}

\vspace{-2mm}
There are two types of asymptotic wave-functions in the bulk gravity theory: the Schwarzschild wave-functions $\Psi_\ell(\nu, t)$ with given frequency  $\nu$ and the Kruskal wave functions $\Phi_\ell(q_-,V)$ and $\Psi_\ell(p_+,U)$ with given ingoing and outgoing Kruskal momentum $q_-$ and $p_+$. All wave functions are assumed to satisfy the wave equation of the particle of mass $m^2_\ell = \ell (\ell -1) $ in AdS units.

We first discuss the Schwarzschild case. We consider an AdS${}_2$ black hole of mass $M$, inverse Hawking temperature $\beta$ and with Schwarzschild coordinates $z$ and $t$. To simplify the expressions we take units in which $\beta=2\pi$.
At the holographic boundary, the light-cone coordinates $u = t+ z$ and $v = t-z$ coincide with the time coordinate $u=v=t$ of the Schwarzian QM. 
In the semiclassical limit, the asymptotic wave functions in this coordinate frame can be obtained by decomposing the boundary-to-boundary propagator as an integral over Schwarzschild frequencies~$\nu$
\bea
\label{gtwoclass}
G_\ell(t_1,t_2) \is
 \left(\frac{1}{2\sinh(\frac{1}{2}(t_{12} - i \epsilon))}\right)^{2\ell} \, = \, \int
 \frac{d\nu}{2\pi}\; \Psi^*_\ell(\nu,t_1)\Psi_
\ell(\nu,t_2),\\[3mm]
\label{wvsch}
& & \Psi_\ell(\nu,t) \, = \,  \frac{e^{\frac{\pi\nu}{2}}
\Gamma\bigl(\ell + i \nu\bigr)}{\sqrt{\Gamma(2\ell)}}\; e^{ -i \nu t}. 
\eea

The asymptotic Kruskal wave functions are bulk-to-boundary propagators sourced on the boundary at the location $(U,V)= (- e^{t}, e^{-t})$ \cite{Shenker:2014cwa, Maldacena:2016upp}. For early times $t_1$, we consider ingoing waves as functions of the ingoing Kruskal coordinate $V$; for late times $t_2$, we consider outgoing waves as functions of the outgoing Kruskal coordinate $U$. Their explicit form is determined by the relations
\bea\label{2ptKruskal}
G_\ell (t_1,t_2) \, = \,\int^{\infty}_0\! \frac{dp_+}{p_+} \, \Psi_{\ell}^*(p_+,U_1)\, \Psi_\ell(p_+,U_2) 
\!\is \! \int^{\infty}_0 \! \frac{dp_-}{p_-}\, \Phi_{\ell}^*(p_-,V_1)\, \Phi_\ell(p_-,V_2), \\[3mm]
\Psi_\ell(p_+,U) = \frac{ \left(ip_+ U\right)^{\ell} }{\sqrt{\Gamma(2\ell)}}\,  {e^{ i p_+ U}}, \qquad \qquad   & & \hspace{-8mm}
\Phi_\ell(p_-,V) \spc = \spc \frac{ \left( ip_- V\right)^{\ell} }{\sqrt{\Gamma(2\ell)}}\, {e^{ i p_- V}},
\eea
where the parameters $p_+$ and $p_-$ are interpreted as bulk null momenta in Kruskal coordinates.
The two types of asymptotic wave functions are related via the unitary basis transformation
\bea\Psi_{\ell}(p_+, U) \!\!\spc \is\!\! \int\! \frac{d\nu}{2\pi} \, p_+^{i\nu} \Psi_\ell(\nu,t) \qquad; \qquad
\Phi_{\ell}(p_-, V) \spc =  \int\!  \frac{ d\nu}{2\pi} p_-^{-i\nu} \Psi^*_\ell(\nu,t) .
\label{basischange}
\eea

\vspace{-2mm}
\subsection{Large $C$ limit of exact two-point function}
\label{semi2pt}
\vspace{-2mm}

We will consider the limit of the two-point function for large $C$ and evaluated at times that might be large but not bigger than $C$. Both integrals appearing in \eqref{2ptExact} are dominated by $k_1 \sim k_2 \gg 1$. With this in mind we will write
\beq
k_1^2=E_1 = M +\omega,~~~k_2^2=E_2 = M,
\eeq
with $\omega \ll M$. 
Using this approximation the two-point function \eqref{2ptExact} becomes 
\bea
\la V_1(0) V_2(t) \ra\is  \int_0^\infty \!\!\! dM \,e^{2 \pi \sqrt{M}-\frac{\beta}{2C} M} \int\! \frac{d\omega}{2\pi}\, e^{ -i\frac{\omega}{2C}t  + \pi \frac{\omega}{2\sqrt{M}}} \, \frac{\Gamma(\ell\pm i\frac{\omega}{2\sqrt{M}})}{2 (2C)^{2\ell}\spc \Gamma(2\ell)}\, (2\sqrt{M})^{2\ell-1}.\
\eea
Rescaling $\omega \to 2C\omega$ and $M \to 2CM $, one immediately sees that in the semiclassical limit the integral over $M$ is dominated by a saddle-point at $M_0= 2\pi^2C/\beta^2 \gg 1$, which is the mass of an AdS$_2$ black hole with inverse Hawking temperature $\beta$. Doing the integral over $M$ reproduces equation \eqref{gtwoclass} if we set $\beta = 2\pi$
\bea
\label{inti}\la V_1(0) V_2(t) \ra \!\! \spc &=&\!\left(\frac{2\pi}\beta\right)^{2\ell-1} \int
\! \spc \frac{d\omega}{2\pi}\, e^{- i t \omega + {\frac \beta  2 \omega}} \frac{\Gamma(\ell\pm i \frac \beta {2\pi} \omega)}{\Gamma(2\ell)} , \nn
&=& \left(\frac{\pi}{\beta \sinh(\frac{\pi}{\beta}(t - i \epsilon))}\right)^{2\ell}.
\eea
This coincides with the semiclassical two-point function derived in the previous section from the AdS wavefunctions.\footnote{If we allow the time difference to be $t\gg C$ then backreaction turns this exponential decay in time into a power-law decay, as found in \cite{altland, MTV}.}


\section{Four-point function} 
\label{sect5}

\vspace{-2.5mm}

In this section we will focus on the Schwarzian four-point function in the large $C$ limit. Our goal is to match the out-of-time ordered (OTO) case with the AdS$_2$ shockwave calculation done in \cite{Maldacena:2016upp}. We will consider ingoing and outgoing matter particles created by local operators $V(t_1)$ and $W(t_2)$ with Lorentzian time difference $t_1-t_2 \gg \beta$. 

For comparison, we first consider the semiclassical limit of the exact time-ordered four-point function \cite{MTV}.  The exact correlation function in the Schwarzian theory takes the following form
\bea
\lb V_1 W_3 V_2 W_4\rb \is \int dk_1^2dk_4^2dk_s^2\sinh(2\pi k_1)\sinh(2\pi k_4)\sinh(2\pi k_s)\nonumber\\[-2mm]\\[-2mm]\nonumber
 \times \ \  & & \hspace{-1cm}
e^{-\frac{1}{2C}\(ik_1^2t_{21}+ik^2_4t_{43}+ik^2_s(-i2\pi-t_{21}+t_{43})\)}\, \frac{\Gamma(\ell_2\pm ik_{4\pm s})\Gamma(\ell_1\pm ik_{1\pm s})}{\Gamma(2\ell_1)\Gamma(2\ell_2)}\,,
\eea
where we took $\beta=2\pi$. We can take the large $C$ limit following the same procedure as used for the two-point function.  A straightforward 
computation shows that the TO four-point function simply factorizes into the product of two-point functions $
G_{\ell_1\ell_2}(t_1,t_2,t_3,t_4) \to G_{\ell_1} (t_1,t_2) \,G_{\ell_2} (t_3,t_4)$.
From the bulk perspective, this result illustrates that the bulk interactions are suppressed in the large $C$ limit. As we will see, this is no longer true for the out-of-time-ordered four point function.

\subsection{Large C limit of OTO four point function}

This OTO amplitude differs from the time-ordered amplitude by an insertion of the $R$-matrix, which contains an additional $u$-integral as shown below. 
The $R$-matrix incorporates the gravitational shockwave interaction. We would like to make this explicit, by comparing the large $C$ limit of the above expression with the Dray-'t Hooft S-matrix.

The result for the OTO four-point function takes the form
\bea \label{intro:OTOC}&  & \qquad \la V_1 W_3 V_2 W_4\ra \, = \,\prod_{i=1,4,s,t}\int d k_i^2\sinh 2\pi k_i\  {\cal A}_{\rm OTO}(k_i,t_i)\\[4mm]
\label{exactoto}
{\cal A}_{\rm OTO}(k_i, t_i)\!\!\spc \is\!\!\spc e^{ - \frac{i}{2C}(  k_1^2t_{31} + k_t^2t_{23} +k_4^2t_{42}+ k_s^2(-i2\pi -t_{41}))} \, \frac{\Gamma(\ell_1\!\!\spc +\!  ik_{1\pm s}) \Gamma(\ell_1\!\!\spc  -\!  ik_{4 \pm t}) \Gamma(\ell_2\!\!\spc - \! \spc ik_{1 \pm t}) \Gamma(\ell_2\!\! \spc + \!  ik_{4 \pm s})}{\Gamma(2\ell_1)\Gamma(2\ell_2)} \nonumber  \\[3mm] & & \hspace{-1.7cm} \times \; \int\limits_{-i\infty}^{i\infty}\!\! \frac{du}{2\pi i} \frac{\raisebox{.5pt}{ $\Gamma(u)\Gamma(u\! -\! 2ik_s)\Gamma(u\! +\!  i k_{1+4-s+t})\Gamma(u\! -\! i k_{s+t-1-4})\Gamma(\ell_1\!  + \! i k_{s-1}\! -\! u)\Gamma(\ell_2\! +\! i k_{s-4}\! -\! u)$}}{\raisebox{-.5pt}{ $\Gamma(u\! +\! \ell_1\!  - \! i k_{s-1})\Gamma(u\! +\! \ell_2\! -\! ik_{s-4})$}},
\nonumber
\eea
where $k_{i+j}$  is shorthand for $k_i + k_j$. The integral over the auxiliary variable $u$ can be done exactly by contour deformation, giving the Wilson function introduced by Groenevelt \cite{groenevelt} (which in turn can be expressed in terms of ${}_4 F_3$ generalized hypergeometric functions). For our study, however, it is more useful to keep the integral expression. 

Similar to what happened with the two-point function, the semiclassical limit of large $C$ is described by a saddle-point where all $k$'s are of order $C$, while its differences are subleading. As we show in appendix \ref{app:R}, by looking at the poles and residues of the integrand, in this limit we only need to take the residue at $u=0$ and $u=ik_{s+t-1-4}$. This gives the following semiclassical expression for the amplitude, in terms of the $k$-variables
\begin{align}
&\la V_1 W_3 V_2 W_4\ra \, =\,  \prod_{i=1,4,s,t}\int d k_i^2\sinh 2\pi k_i\; e^{ - \frac{1}{2C}\(  ik_1^2(t_3-t_1) + ik_t^2(t_2 -t_3) +ik_4^2(t_4-t_2)+ ik_s^2(-i2\pi-t_4 +t_1)\)} \nonumber
\\[1mm]
&\times\! \frac{\Gamma(\ell_1\!+\!ik_{s\pm 1}) \Gamma(\ell_1 \!-\! ik_{4 \pm t}) \Gamma(\ell_2\! -\! ik_{1 \pm t}) \Gamma(\ell_2 \!+ \!ik_{s \pm 4}) \Gamma(-\! 2ik_s)\Gamma(i k_{-s+t+1+4})\Gamma(i k_{-s-t+1+4})}{\Gamma(2\ell_1)\Gamma(2\ell_2)} \nonumber
\\[3mm]
&\qquad\qquad\qquad \qquad \qquad  +\  \    (k_1,k_4, k_s, k_t) \, \rightarrow \, (-k_4,-k_1,-k_t,-k_s)
\,.
\end{align}
The first and the second term in the right hand side of the previous equation come from the pole at $u=0$ and $u=ik_{s+t-1-4}$ respectively. They are related by taking $k_s\rightarrow -k_{t}$, $k_t\rightarrow -k_{s}$, $k_1\rightarrow -k_{4}$, $k_4\rightarrow-k_1$ as indicated. In the limit of large time difference between the $V$ and $W$ operators the contribution from the second pole is negligible, and the $u=0$ pole dominates. We will accept this for now and come back to the role of this extra term at the end of this section. 

Following the procedure we outlined for the analysis of the two point function, we rewrite the integrals using the following variables
\beq
k_s^2=M,\quad k_1^2=M+\omega_1,\quad k_4^2=M+\omega_4,\quad k_t^2=M+\omega_4+\omega_2\,.
\eeq
In the $\omega\ll M$ limit, the OTO correlation function becomes
\begin{align}
&\langle V_1 W_3 V_2 W_4\rangle\; \simeq\; \int d M e^{2\pi\sqrt{M}-\frac{\pi}{C} M} (2\sqrt{M})^{2\ell_1+2\ell_2+i\nu_1-i\nu_2-4}\qquad \qquad
\\[1mm]
& \qquad \  \ \prod_{i=1}^4\int\! \frac{d\nu_i}{2\pi} \; e^{i\nu_1 t_1-i\nu_2 t_2 +i\nu_3 t_3 -i\nu_4 t_4+\frac{\pi}{2}(\nu_2+\nu_1+2\nu_4)}\, 2\pi \delta(\nu_1+\nu_3-\nu_2-\nu_4)\nonumber
\\[1mm]
&\; \qquad \qquad \ \ \  \times\; \frac{\Gamma(\ell_1-i\nu_1)\Gamma(\ell_1+i\nu_2)\Gamma(\ell_2+i\nu_3)\Gamma(\ell_2-i\nu_4)\Gamma(i\nu_1-i\nu_2)}{\Gamma(2\ell_1)\Gamma(2\ell_2)}  \nonumber
\, ,
\end{align}
where we define  $\nu=\frac{\omega}{2\sqrt{M}}$, and rescaled $t_i$ accordingly, to simplify the notation. Just as before, the  $M$ integral is dominated by the saddle point at $M_0 = 2\pi^2C / \beta^2$. The above result then manifestly matches the flat space shockwave calculation in section \ref{sect2} for a massless (conformally coupled in 2D) particle with $\ell=1$. It can, equivalently, be read as being composed of bulk-to-boundary propagators of the type \eqref{wvsch} and the $\mathcal{S}$-matrix \eqref{smatf}.

We conclude with a comment about the contribution in the $u$-integral coming from the pole at $u=ik_{s+t-1-4}$. Since the form of this term is very similar to the shockwave integrand, one  can perform the integral exactly and check that it vanishes in the large time limit. Nevertheless, there is an important reason why this term is present. In this section we have been implicitly focusing on the  OTO four-point function $\lb V(0) W(t) V(0) W(t)\rb$. If one had chosen $t$ to be large but negative, we would instead have  been computing $\lb V(t) W(0) V(t) W(0)\rb$. The expression \eqref{eq:MSYsw} we write down below for the shockwave answer breaks $t \to -t$ symmetry and would not be valid in this case. If we had computed $\lb V(t) W(0) V(t) W(0)\rb$ we would have found that the roles of the $u$-poles are reversed. Namely, the $u=0$ pole would be negligible, and the $u=ik_{s+t-1-4}$ pole would reproduce the shockwave calculation. This structure fits nicely into what we would expect from a non-perturbative formula taken in the semiclassical limit.

The analysis of this and the previous section can be generalized to arbitrary higher-point OTO correlation functions. We present a few examples in appendix \ref{secthigh}.

\subsection{OTO four point function as a Virasoro conformal block}
\label{sectblock}

The OTO four-point function can be expressed in terms of this bulk $\mathcal{S}$-matrix  and the asymptotic wave functions. The integrals can be computed explicitly, with the result \cite{Maldacena:2016upp} \footnote{Even though it is not obvious from this expression one can verify using the properties of the hypergeometric function $U(a,b,z)$ that the right hand side is invariant under $\ell_1 \leftrightarrow \ell_2$.}
\bea\label{eq:MSYsw}
\frac{\lb V_1 W_3 V_2 W_4\rb}{\lb V_1 V_2 \rb \lb W_3 W_4\rb}  \; \is z^{-2 \ell_1} U(2 \ell_1, 1+2\ell_1-2\ell_2 , 1/z),
\eea
where we define the cross-ratio 
\bea
\label{xratio}
z \is \frac{i\beta}{16\pi C} \frac{e^{\pi(t_3 + t_4 - t_1 -t_2)/\beta}}{\sinh \frac{\pi t_{12}}{\beta} \sinh \frac{\pi t_{34}}{\beta}}.
\eea
If we make the choice of the insertions times similar to \cite{Maldacena:2016upp}, explicitly $t_1 = - i\frac{\beta}{2}$, $t_2=0$, $t_3 = t - i \frac{\beta}{4}$ and $t_4 = t + i\frac{\beta}{4}$, then the cross-ratio becomes $z = \frac{\beta}{16\pi C} e^{\frac{2\pi}{\beta}t} $. The shockwave calculation is valid for $t>0$ large with this combination $z$ held fixed. 

In the previous section we got this result from our exact formulas derived from 2d Liouville CFT between ZZ-branes. The purpose of this section is to rederive the semiclassical limit directly from the 2d picture without having to go through the details of the exact expressions. Instead of taking large $C$ with fixed $\beta$ we will consider units in which $C=1/2$. Since the dimensionless coupling is $2\pi C/\beta$ the semiclassical limit is equivalent to taking $\beta \to 0$ in these units. 

In the 2d picture, the inverse temperature $\beta$ of the Schwarzian gives the distance between the ZZ branes. Taking $\beta \to 0$ in the Schwarzian means sending the distance between the ZZ-branes to zero faster than the size of the circle in the extra dimension. Namely, $\beta$ goes to $0$ faster than $q \to 1$, where $q=e^{2 \pi i \tau}$ denoted the $q$-modulus of the 2d annulus. In this limit the Schwarzian becomes equivalent to Liouville between two infinite ZZ-branes, namely on a strip of width $\beta$ instead of an annulus. 

The upshot of the previous argument is that we can reproduce the semiclassical Schwarzian correlators from local operators between two infinite ZZ-branes. The Liouville one-point function, which corresponds to the Schwarzian two-point function, is easy to compute exactly from the 2d CFT perspective, since the system can be mapped to the upper-half-plane by a conformal transformation. The answer immediately has the required form 
\beq
\lb V \rb_{\rm strip} = \left(\frac{\pi }{\beta \sin \frac{\pi \tau}{\beta}} \right)^{2\ell},
\eeq
where $\ell$ corresponds to the conformal dimension of the Liouville operator. This can be related to the real time answer \eqref{gtwoclass} by analytic continuation.

Now we compute the Liouville 2pt function/Schwarzian 4pt function using this approach. Again, we can map the infinite strip to the upper-half plane, and we take the positions of the two local vertex operator insertions to be $z_1$ and $z_3$, while the images of these operators will be denoted by $z_2$ and $z_4$ (even though they should strictly be given by $z_1^*$ and $z_3^*$ we will allow them to be generic). The two-point function can be written in two equivalent ways. First, we can take the OPE between the two insertions and between the two images, obtaining
\bea
\frac{\la V_1  V_2 W_3W_4 \ra}{\la V_1V_2\ra\la W_3 W_4\ra} \! \is\!\! \int \! dP\, \Psi_{\rm ZZ}(P)\, C_{VWP} \, \mathcal{F} \Big( {}^{V}_{W} \; {}^{V}_{W} , P, \eta \Big), \ \ \ \
\eea
where  $\eta=\frac{z_{13}z_{24}}{z_{14}z_{23}}$ is the cross-ratio, $\Psi_{\rm ZZ}$ is the ZZ-brane wavefunction, $C_{VWP}$ represents the Liouville OPE coefficient between the operators $V$, $W$ and an intermediate channel operator with Liouville momentum $P$. $ \mathcal{F} (P, \eta )$ denotes the conformal block in this channel. Another representation of this correlation function can be obtained by performing the OPE between an operator and its image. In this case it was shown that only the vacuum block appears (see section 6 of \cite{Zamolodchikov:2001ah} and also \cite{LeFloch:2017lbt} for a different perspective on this result).
Defining the new cross ratio via $x=1-\eta$ and using the exponential map $z_i =e^{\frac{2\pi}{\beta} \tau_i}$, the ZZ identity gives 
\beq
\frac{\lb V_1 V_2 W_3 W_4\rb }{\lb V_1 V_2\rb \lb W_3 W_4\rb} = x^{2\Delta_V}  \mathcal{F} \Big( {}^{V}_{V} \; {}^{W}_{W} , {\rm vac},x\Big),~~~~x= -\frac{\sinh \frac{\pi t_{12}}{\beta} ~\sinh \frac{\pi t_{34}}{\beta} }{\sinh \frac{\pi t_{32}}{\beta} ~\sinh \frac{\pi t_{41}}{\beta} }.
\eeq
For fixed $t_1,\ldots, t_4 \sim \mathcal{O}(1)$ and $c\to \infty$, the cross-ratio is finite and the vacuum block becomes trivial, implying that $\lb V_1V_2W_3W_4\rb \sim \lb V_1V_2\rb \lb W_3W_4 \rb$.  For the time-ordered four-point function this is the final answer.

The out-of-time ordered four-point function is equal to the vacuum block evaluated on the second sheet. 
It turns out this indeed exactly reproduces the shockwave calculation. 
The vacuum block on the second sheet is found by performing a monodromy operation on the block. As observed in \cite{Kaplan}, this monodromy remains non-trivial in the combined $x\to 0$ and $c\to \infty$ limit, with the product $c \cdot x$ is held fixed and finite. The exact formula for the identity block in this limit was found to be \cite{Kaplan}
\bea
\frac{\lb V_1 W_3 V_2 W_4\rb }{\lb V_1 V_2\rb \lb W_3 W_4\rb}&=&\lim_{c\to \infty, cx~{\rm fixed}} x^{2\Delta_V}  \mathcal{F}_{2^{nd}{\rm sheet}} \Big( {}^{V}_{V} \; {}^{W}_{W} , {\rm vac},x\Big) \nonumber \\[-2mm]\\[-2mm] \nonumber
&=& z^{-2\ell_1} U(2\ell_1,1+2\ell_1-2\ell_2,1/z),
\ea
where the right hand side involves the cross ratio $z$ defined in equation \eqref{xratio}.
Here we used the precise relation between the Virasoro central charge $c$ and the Schwarzian coupling $2\pi C/\beta$.  This matches exactly with the shockwave calculation in equation \eqref{eq:MSYsw}.

\section{Large $\ell$ limit of the exact correlation function}
\label{sectell}

\vspace{-2mm}

In this section we study the limit in which we simultaneously take the mass $\ell$ of the insertions to be heavy in combination with $C$ large, keeping $\ell/C$ finite. We will take the relevant limit directly in the Euclidean regime of our expression \eqref{2ptExact} (and compare that to a direct solution of the Schwarzian equation of motion in appendix \ref{sectheavy}). This regime was explored from a geometrical perspective in \cite{Gu:2017njx}. 

\subsection{Two point function}\label{2ptspsec}
Consider the two-point function for a Schwarzian bilocal operator, where we write the operator in the action:
\bea
\label{semipath}
G_\ell(\tau)\! \is \! \int \left[\mathcal{D}f\right]  \exp(-S)\qquad {\rm with} \qquad 
S \, = \,  -C \int dt \left\{F,\tau \right\} - \ell \ln \frac{\dot{F}_1\dot{F}_2}{(F_1-F_2)^2}.
\eea
Here $F = \frac{\beta}{\pi}\tan\left(\frac{\pi}{\beta}f\right)$ where $f$ maps the interval $I_\beta$ monotonically into itself: $f(\tau +\beta) = f(\tau)+ \beta$.
Within the semiclassical regime, we set the weight of the bilocal operator to scale as $\ell \sim \mathcal{O}(C)$ with $C \to +\infty$, such that both terms in the action are equally important.
In this section we would like to make a connection between the large $\ell$ limit, with $\ell/C$ fixed, of the exact two-point function and extract a practical geometric representation along the lines of \cite{Gu:2017njx}\cite{Kourkoulou:2017zaj}.

The exact two-point function \eqref{2ptExact}, in Euclidean signature, can be simplified in this semi-classical regime ($C\gg \tau, \beta-\tau$) by using Stirling's formula to
\bea
\label{ex2ptlim}
\la V_1(0) V_2(\tau) \ra \!\! & \approx &\! \int\! dk_1^2 dk_2^2 \, e^{-\frac{\tau k_1^2}{2C} -\frac{(\beta-\tau)k_2^2}{2C}}e^{2\pi (k_1+k_2)}\((k_1-k_2)^2+\ell^2\)^{\ell-\frac 1 2}\((k_1+k_2)^2+\ell^2\)^{\ell-\frac 1 2} \nonumber \\[2.5mm]
& &\qquad \quad  \times \frac{1}{(2C)^{2\ell}(2\ell)^{2\ell}}\; \, e^{-\mbox{\footnotesize $2(k_1+k_2)\spc \text{arctan}\bigl(\frac{k_1\! +\nspc k_2}{\ell}\bigr)-2(k_1\! -\nspc k_2)\spc \text{arctan}\bigl(\frac{k_1-k_2}{\ell}\bigr)$}}. \nonumber
\eea
These momentum integrals are dominated by their saddle points. The saddle point equations are
\bea
\label{finitetsaddle}
\frac{\tau k_1}{C} + 2\spc \text{arctan}\Bigl(\frac{k_1+k_2}{\ell}\Bigr) + 2\spc \text{arctan}\Bigl(\frac{k_1-k_2}{\ell}\Bigr) &= 2\pi, \nonumber \\[-2mm]\\[-2mm]\nonumber
\frac{(\beta-\tau) k_2}{C} + 2\spc \text{arctan}\Bigl(\frac{k_1+k_2}{\ell}\Bigr) - 2\spc \text{arctan}\Bigl(\frac{k_1-k_2}{\ell}\Bigr) &= 2\pi,
\eea
which demonstrates that $k_i \sim \mathcal{O}(C)$. Performing the saddle point gaussian integral yields the semi-classical result 
\bea
\label{semiclass}
G^{\beta}_\ell(\tau)\!\! &  \simeq & \!\! e^{-S_0} \sqrt{\frac{\pi}{\text{det}\mathbf{A}}},
\eea
with classical ``action''\footnote{Here and in what follows, we denote the solution of the saddle equations by $k_i$ as well.}
\bea
\label{sadintT}
S_{\rm class.} \is \frac{\tau k_1^{2}}{2C} + \frac{(\beta-\tau) k_2^{2}}{2C} + \ell \ln\left[\frac{((k_1+k_2)^2+\ell^2)((k_1-k_2)^2+\ell^2)}{(4C\ell)^2}\right],
\eea
and fluctuation determinant
\bea
\text{det} \mathbf{A} \is \frac{\tau(\beta-\tau)}{4C^2} + \frac{\frac{\beta \ell}{C}(\ell^2+k_1^{2}+k_2^{2}) + 4\ell^2}{((k_1 + k_2)^2 + \ell^2)((k_1 - k_2)^2 + \ell^2)}.
\eea
We would like to give these formulas a more evident geometrical interpretation, at least for the classical action. Nevertheless this variational problem does not have enough variables to be compared with the geometric one, as we explain below. Therefore we need a different representation of the two-point function.

We again start with the exact two-point function \eqref{2ptExact}.
To take the large $\ell$ limit of this formula in a way that will help make the geometric interpretation more transparent, we first rewrite the product of gamma functions in the exact two-point function \eqref{2ptExact} using the following formula\footnote{We are still using our short-hand notation $\Gamma(a\pm b) = \Gamma(a+b)\Gamma(a-b)$.}
\beq
\frac{\Gamma(\ell \pm i(k_1 \pm k_2))}{\Gamma(2\ell)^2}=\int \frac{dy_1}{\pi}\frac{dy_2}{\pi} \frac{e^{i 2(k_1+k_2) y_1+i 2 (k_1-k_2) y_2}}{(4 \cosh y_1 \cosh y_2)^{2\ell}}.
\eeq
This allows to write the two-point function \eqref{2ptExact} as an integral over $k_i$ and $y_i$. In the large $C$ and large $\ell$ limit we can approximate the density of states by a simple exponential and rewrite the integrand as
\beq
\la V_1(0) V_2(\tau) \ra  = \prod_{i=1,2} \int \! dy_i dk_i \; \;  \exp\Big( - S_0 - S(k_1,k_2,y_1,y_2) \Big),
\eeq
where $S_0$ is independent of the variational parameters $k$'s and $y$'s, and the classical action in the exponent is given by
\beq
S=-(2 \pi +2 i y_1+2 i y_2 )k_1-(2\pi +2 i y_1 - 2 i y_2)k_2+\frac{k_1^2}{2C}\tau + \frac{k_2^2}{2C}(\beta-\tau)+ \ell \log (\cosh(y_1)\cosh(y_2))^2,
\eeq
where we ignore terms that are subleading in this limit. In a large $C$ limit with $\ell/C$ fixed this is dominated by a saddle-point at real $k$'s and complex $y$'s. To make the physics of this expression more transparent we change variables as 
\bea
2\pi +2 i y_1+2 i y_2 =\theta_1, \quad
&& \quad 2\pi +2 i y_1 - 2 i y_2= \theta_2.
\ea
The action then simplifies to
\beq\label{actionlargeL}
S(k_i,\theta_i)=\frac{k_1^2}{2C}\tau +\frac{k_2^2}{2C}(\beta-\tau) - \theta_1 k_1- \theta_2 k_2 + \ell \spc \log \left(\frac{ \cos \frac{\theta_1}{2}+\cos \frac{\theta_2}{2}}{2}\right)^2.
\eeq
The saddle-point equations of this action become 
\bea
 k_1=-\frac{\ell}{2}\left( \tan \frac{\theta_1+\theta_2}{4} +  \tan \frac{\theta_1-\theta_2}{4}\right), \qquad &&\theta_1 =\frac{k_1}{C} \tau, \nn
k_2 = -\frac{\ell}{2}\left( \tan \frac{\theta_1+\theta_2}{4} -  \tan \frac{\theta_1-\theta_2}{4}\right),\qquad
&&\theta_2 = \frac{k_2}{C} (\beta -\tau) .
\ea
After solving for $\theta_1$ and $\theta_2$, one recovers the saddle-point formulas of \eqref{finitetsaddle}.\footnote{We thank Z. Yang for discussions on these variables.} In particular, the saddle-point gives $k\sim \mathcal{O}(C)$, $\theta \sim \mathcal{O}(1)$ and $S \sim \mathcal{O}(C)$.

\begin{figure}
\begin{center}
\begin{tikzpicture}[scale=0.6]
\draw[thick,fill,gray, opacity=0.15] (0,0) circle (4);
\draw[blue,line width=1] (-1.9,-1.9) arc (240:26:2.74);
\draw[blue,line width=1] (-1.9,-1.9) arc (26:240:-2.74);
\draw[dashed] (-1.9,-1.9) -- (1.9,1.7);
\draw[thick] (0,0) circle (4);
\draw[fill] (-1.9,-1.9) circle (0.11);
\draw[fill] (1.9,1.7) circle (0.11);
\draw (-1.3,1.3) node {\large $A_1,L_1$}; 
\draw (1.3,-1.3) node {\large $A_2,L_2$}; 
\draw (-2.4,-2.2) node {\Large $X$}; 
\draw (2.4,2) node {\Large $Y$}; 
\end{tikzpicture}
\end{center}
\caption{Geometric minimization problem. The gray circle is euclidean AdS$_2$. The blue line is the cut-off boundary of AdS. $X$ and $Y$ correspond to the insertions of the two-point function. We separate the boundary in two arcs of length $L_1$, $L_2$ and enclosing area $A_1$,$A_2$.}
\label{fig:geoint}
\end{figure}
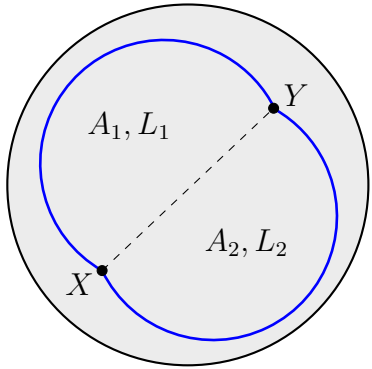

Following \cite{Gu:2017njx} this can be given a geometric meaning. In that work the authors show that the calculation of the two-point function is equivalent to the geometric problem of minimizing an action proportional to the sum of the area enclosed by the boundary curve and an extra term 
\beq
\ell \spc \log \cosh D,
\eeq
where $D$ is the geodesic distance between the insertions of the two-point function.

The minimization can be done in steps. First one can minimize both halves of the boundary curve independently and get two arcs of a circle.  We show this configuration in figure \ref{fig:geoint}. Then the minimization is done with respect to the opening angle and its area. This is manifest in our formula \eqref{actionlargeL}. The variables $\theta_1$ and $\theta_2$ correspond to the opening angle of both circles while $k_1$ and $k_2$ are inversely proportional to the radius of each arc. Moreover the first four terms in \eqref{actionlargeL} are proportional to the total area inside the boundary curve while the term proportional to $\ell$ corresponds to the geodesic length between the boundary insertions. 

For concreteness one can check this with an example. If $\ell/C$ is fixed but much smaller than one then the solutions is $k_1 \sim k_2 =2\pi C /\beta$ and $\theta_1=2\pi \frac{\tau}{\beta}$, $\theta_2 = 2 \pi \frac{\beta-\tau}{\beta}$. This is consistent with the two-point function that do not backreact the geometry and simply computes the renormalized geodesic distance between the two points in AdS. On the other hand for large $\ell/C$, the solution is $k_1 \sim 2\pi C/\tau$, $ k_2 \sim 2\pi C/ (\beta-\tau)$ and $\theta_1 \sim \theta_2 \sim 2\pi$. In this limit the points $X$ and $Y$ become close and the boundary turns into two full circles touching at a point, as in figure 7 of \cite{Kourkoulou:2017zaj}.

All of this matches with a classical solution of the Schwarzian equations of motion, sourced by the heavy bilocal operator. We present this computation in appendix \ref{sectheavy}.

\subsection{Application: Eigenstate Thermalization Hypothesis }
As a final application, we will study the Eigenstate Thermalization Hypothesis (ETH) using the exact expressions from \cite{MTV} (for discussions in the context of the SYK model see \cite{Kourkoulou:2017zaj}\cite{Eberlein:2017wah}\cite{Sonner:2017hxc}). For this purpose we will evaluate the two-point function of a light operator $V$ of dimension $\ell$ between energy eigenstates  
\beq
\lb E | V (t) V(0) | E\rb \approx \text{Tr}\left[ e^{-\beta(E) H} V(t) V(0)\right],
\eeq
where the energy $E$ of the eigenstate is large and will be taken of order $C$. We will check this statement in the Schwarzian theory where $\beta(E)$ is a function (given below) of the eigenstate's energy. We will achieve this by taking the semiclassical limit of non-perturbative formulas. 

Schwarzian energy eigenstates are labeled by a parameter $k$, which is real and positive, with energy $E(k)=k^2/2C$ and density of states $\rho(k)= 2k \sinh 2 \pi k$. The expectation value we need is given by
\beq
\lb E | V (t_1) V(t_2) | E\rb = \lb k(E) | \mathcal{O}_\ell (t_1,t_2) | k(E)\rb,
\eeq
in terms of the Schwarzian bi-local field $\mathcal{O}_\ell(t_1,t_2)$ and with $k(E) = \sqrt{2 C E}$. An exact expression for this quantity can be easily extracted from the exact two point function at finite temperature \eqref{2ptExact}. The final answer is 
\begin{equation}\label{expvalETH}
\lb E | V(t_1) V(t_2) | E\rb=\int dp^2\sinh(2\pi p) e^{-i\frac{1}{2C}(p^2-k^2)t_{12}}\frac{\Gamma(\ell \pm i(p\pm k))}{\Gamma(2\ell )}\,,
\end{equation}
where again we use $k=\sqrt{2C E}$. It is important to note that even though we extracted this from the thermal two-point function, the right hand side of \eqref{expvalETH} is a zero-temperature expectation value between energy eigenstates. To study ETH we take $\ell \ll C$ and $k(E)\sim C$ with large $C$. Similar to the situation in section \ref{semi2pt}, the integral is dominated by $p \approx k$. Therefore we can rewrite the integral using the following variables
\begin{equation}
k^2=M ,\quad p^2=M +\omega \, ,
\end{equation}
where we follow the same notation as in section \ref{semi2pt}. Then in the same way we obtained the semiclassical limit of the two-point function, we find the integrals are dominated by $\omega \sim \mathcal{O}(C) \ll M \sim \mathcal{O}(C^2)$. Applying the same approximations used in section \ref{semi2pt} for the gamma functions, the two-point function on an energy eigenstate becomes
\bea
\lb E | V(t_1) V(t_2) | E\rb \approx \int d\omega ~ e^{-i\frac{\omega }{2C}t_{12}+\pi\frac{\omega}{2\sqrt{M}}}\frac{\Gamma(\ell \pm i\frac{\omega}{2\sqrt{M}})}{\Gamma(2\ell)}(2\sqrt{M})^{2\ell-1}\,.
\ea
This is precisely the Fourier transform of the thermal two-point function in terms of Lorentzian times $t_1$ and $t_2$ with the effective inverse temperature $\beta=2 \pi C/ \sqrt{M}$. Then performing the integral over $\omega$ gives
\beq
\lb E | V(t_1) V(t_2) | E\rb \approx \left( \frac{\pi}{\beta(E) \sinh \frac{\pi}{\beta(E)} t_{12} }\right)^{2 \ell}= {\rm Tr}[e^{-\beta(E) H} V(t_1) V(t_2)],
\eeq
where the effective inverse temperature involved in the right hand side is a given function of the eigenstate energy 
\beq
\beta (E) = 2 \pi \sqrt{\frac{C}{2E}}.
\eeq
This shows how the two-point function of light operators thermalizes in an eigenstate of high energy. It is important to notice that we needed to take $E$ large and of order $C$. For example, had we taken large $C$ and $E\sim \mathcal{O}(1)$, the integral appearing in \eqref{expvalETH} would have no simplification, the expectation value would look very different from thermal and the eigenstate would not have thermalized.

Finally we will compute the semiclassical limit of the heavy-light four-point function at zero temperature. We take $W$ of dimension $\ell_2$ to be heavy while $V$ is a light operator of dimension $\ell_1$. Using the methods from \cite{MTV}, the relevant correlator is
$$
\hspace{-3mm}\langle W_1V_2V_3W_4\rangle=\int dk_1^2dk_2^2\sinh(2\pi k_1)\sinh(2\pi k_2)e^{i\frac{k_1^2}{2C}(t_{21}+t_{43})+i\frac{k_2^2}{2C}t_{32}}\frac{\Gamma(\ell_1\pm ik_{1\pm 2})\Gamma(\ell_2\pm ik_1)^2}{\Gamma(2\ell_1)\Gamma(2\ell_2)}\,.
$$
We will take large $C$ with the heavy operator $W$ satisfying $\ell_2\sim C$ while for the light one $V$ instead $\ell_1\ll C$. We have put the heavy operators at Lorentzian times $t_1$ and $t_4$ for now, below we will take the limit $t_{14} \to \infty$. We can rewrite the integral using the following variables $k_2^2=M$ and $ k_1^2=M+\omega$ with $\omega \ll M$. Applying the appropriate approximation used in the semiclassical limit and the large $\ell$ limit for two-point function, the four-point function becomes
\begin{equation}
\begin{aligned}
\langle W_1V_2V_3W_4\rangle\approx&\int dM \frac{(\ell_2^2+M)^{2\ell_2-1}}{(2\ell_2)^{2\ell_2}}e^{2\pi \sqrt{M}-\frac{iM}{2C}t_{41}-4\sqrt{M}\arctan\frac{\sqrt{M}}{\ell_2}}
\\
&\qquad\quad\times\int d\omega e^{-\frac{i\omega }{2C}t_{32}+\pi\frac{\omega}{2\sqrt{M}}}\frac{\Gamma(\ell_1\pm i\frac{\omega}{2\sqrt{M}})}{\Gamma(2\ell_1)}(2\sqrt{M})^{2\ell_1-1}\,.
\end{aligned}
\end{equation}
The first integral is dominated by the saddle point, similar to the situation in section \ref{2ptspsec}. The saddle-point equation fixes the value of $M$ in terms of $\ell_2$. In the large $t_{41}\to \infty$ limit the solution is $\sqrt{M} = i \ell_2$. After substituting the saddle point value of $M$, the second integral can be identified as the thermal two-point function 
\beq
\langle W_1V_2V_3W_4\rangle_{\beta \to \infty}=\left( \frac{\pi}{\alpha(\ell_2) \sin \frac{\pi}{\alpha(\ell_2)} t_{23} }\right)^{2 \ell_1}, \quad \alpha(\ell) = \frac{2\pi C}{\ell}.
\eeq
In analogy with the situation in CFT$_2$/AdS$_3$ we interpret this result in the following way. The heavy operator $W$ of dimension $\ell_2$ creates a conical defect geometry in the bulk with parameter $\alpha(\ell_2)$.

It is interesting to compare the results of this section with the situation in 2d CFT. In the latter, the state-operator correspondence relates the energy eigenstate calculation to a heavy-light correlator, and one can see a transition between the creation of a conical defect and a BTZ black hole as we increase the scaling-dimension/energy. The situation in the Schwarzian theory is different due to the lack of state-operator correspondence. High energy eigenstates $|k\rb$ (which do thermalize) are not related to heavy operators of dimension $\ell$ (which create conical defects instead). Looking at the derivation of the Schwarzian action from 2d Liouville \cite{MTV} one can trace this back to the fact that Liouville itself does not have a state-operator correspondence (see for example \cite{Seiberg:1990eb}).

\section{Concluding Remarks}
\label{conclusion}
\def\J{\mathcal{J}}
In this paper we have studied in detail the correlators in the Schwarzian / Jackiw-Teitelboim theory. Our analysis was based on the exact formulas found in \cite{MTV} for these quantities. In particular, we have verified (in the semiclassical limit) the proposal put forward in \cite{MTV} that the R-matrix, given by the $6j$-symbols of $SL(2,\mathbb{R})$, controls the out-of-time-ordered correlators of the Schwarzian theory and also correspond to the gravitational S-matrix in the 2d Jackiw-Teitelboim gravity theory. This resonates with the ideas put forward in \cite{JMV} for the case of 3d gravity. 

As a side comment, in this paper we have focused on the semiclassical limit of large $2 \pi C/\beta$ from the perspective of the non-perturbative expressions. We have also taken time differences between operators insertions to be large but smaller than $C$. When $t \gg C$ quantum effects become important and correlation functions, even OTO, go to power laws with different exponents \cite{altland} \cite{MTV}. It would be interesting to understand this cross-over from a bulk perspective.

We would like to conclude by describing an interesting open problem that we leave for future work. In this paper we have analyzed different semiclassical limits of the exact correlators of the Schwarzian theory \cite{MTV}. These results have been obtained as a certain limit of 2d Liouville CFT. In this section we want to raise some points that give a new perspective on this approach. 

 The Schwarzian theory arises as the low energy limit of holographic quantum mechanical models \cite{KitaevTalks}. The main example is the SYK model of $N$ Majorana fermions $\psi_i$ with Hamiltonian 
\beq
H= i^{q/2} \sum j_{i_1\ldots i_q} \psi_{i_1}\ldots \psi_{i_q},
\eeq 
where the disorder average over couplings $j$ is described by 
\beq
\lb j^2_{1\ldots q}\rb = \J^2 \frac{2^{q-1}(q-1)!}{q N^{q-1}}.
\eeq 
As explained in \cite{KitaevTalks} one can reformulate this theory and go from a path integral over $\psi$ and $j$ to a mean field formulation with fundamental fields $G(\tau_1,\tau_2)$ and $\Sigma(\tau_1,\tau_2)$. The former is identified with 
\beq
G(\tau_1,\tau_2) \equiv \frac{1}{N} \sum_i \lb \psi_i (\tau_1)\psi_i(\tau_2)\rb,
\eeq
 and the latter with the self-energy. Fermion correlators can then be replaced by correlators of this bilocal mean field $G(\tau_1,\tau_2)$, integrated over with a semiclassical action \cite{Maldacena:2016hyu}
\beq
-S_E/N = \frac{1}{2} {\rm Tr} \log{(\partial_\tau -\Sigma)} - \frac{1}{2} \int d\tau_1d\tau_2 \left[\Sigma(\tau_1,\tau_2)G(\tau_1,\tau_2) - \frac{\J^2}{q^2} G(\tau_1,\tau_2)^q \right].
\eeq
Analyzing the saddle point equations associated to this action one can find that in the strong coupling limit of large $\beta \J$ the two-point function becomes $G(\tau_1,\tau_2)\sim  |\tau_{12}|^{-2\Delta}$ with scaling dimension $\Delta=1/q$. We will focus now on the large $q$ limit. This means we can approximate the bilocal field in the following way up to $1/q$ corrections 
\beq
G(\tau_1,\tau_2) = \frac{{\rm sgn}(\tau_{12})}{2} e^{- \Delta g(\tau_1,\tau_2)}= \frac{{\rm sgn}(\tau_{12})}{2} \left(1+\frac{1}{q} g(\tau_1,\tau_2)\right),
\eeq
and study the dynamics of $g(\tau_1,\tau_2)$. On-shell the self-energy is also given in terms of $g(\tau_1,\tau_2)$ as $\Sigma \sim \mathcal{J}^2e^{g(\tau_1,\tau_2)}/q^2$. Since we are interested in fermion correlators that can be obtained from correlators of the bilocal field $g(\tau_1,\tau_2)$ one can integrate first over $\Sigma$. This can be done in the large $q$ limit to obtain an effective action for $g$. This was done in \cite{Maldacena:2016hyu,Cotler:2016fpe, Maldacena:2018lmt} giving 
\beq
S_{\rm eff} = \frac{N}{8q^2} \int d\tau_1d\tau_2 \left[\partial_{\tau_1} g \partial_{\tau_2} g - 4 \J^2 \exp{g(\tau_1,\tau_2)}  \right].
\eeq
It was also noted that this is precisely the Liouville action for $g(\tau_1,\tau_2)$. This bilocal action from the point of view of the original quantum mechanical system becomes local in the two dimensional kinematic space $(\tau_1,\tau_2)$. These two parameters behave like null coordinates in the 2d space $(x^0, x^1)$ such that 
$z = \tau_1 = -x^0 + x^1,$ $\bar{z} = \tau_2=x^0 + x^1$
and $g(\tau_1,\tau_2) \to g(z,\bar{z})$. Then we can use this relabeling to write the action as a 2d theory for a scalar field $g$ 
\bea
S_{\rm eff} \is \frac{N}{8q^2} \int d^2z\left[ \partial g \bar{\partial} g - 4 \J^2 \exp{g}  \right].
\eea
We should compare this with standard Liouville CFT.  Liouville theory with a cosmological constant $\mu$ and  central charge as $c=1+6(b+1/b)^2$ is described by the action
$S_{\rm L} = \frac{1}{4\pi}  \int d^2z\left[ \partial \phi \bar{\partial}\phi  + 4 \pi \mu e^{2 b \phi} \right].$
We will take the limit $b\to0$ and therefore $c=6/b^2$. To make contact with the effective SYK-model action we change variables $2b \phi \to g.$
This turns the Liouville action into $S_L =\frac{1}{16\pi b^2}  \int d^2z\left[ \partial g \bar{\partial}g  + \hat{\mu} e^{g} \right],$
where $\hat{\mu} = 16 \pi \mu b^2$ is finite in the $b\to 0$ limit. This allows us to identify the relevant parameters of the 2d CFT with the SYK mean field action. The renormalized cosmological constant is $\hat{\mu}\sim \J^2$, and the central charge is given by
\bea
c\is \frac{12 \pi N}{q^2}.
\eea
Moreover, the bilocal field $G \sim e^{ \Delta g}$ precisely corresponds to a Liouville primary operator $V_\Delta = e^{2 b \Delta \phi}$, which in the small $b$ limit has conformal dimension $\Delta$ (for the specific value of small $\Delta = 1/q$). This allows us to relate the general  correlators via
\bea
\lb G(\tau_1,\tau_2) \ldots G(\tau_{n-1},\tau_n) \rb_{\rm SYK} \is \lb V_{\Delta} (z=\tau_1,\bar{z}=\tau_2)\ldots V_{\Delta} (z_{n/2}=\tau_{n-1},\bar{z}_{n/2}=\tau_n) \rb_{\rm Liouville},
\nonumber
\eea
as anticipated in \cite{MTV}.

There is a subtlety in the above discussion regarding boundary conditions. In the SYK context the right boundary conditions are given by $g(\tau_1,\tau_2)\to 0$ as $\tau_{12}\to 0$. This is consistent with the UV of the theory being described by the free fermion model. With this boundary conditions the Liouville analysis would reproduce the full SYK correlators. Nevertheless this boundary condition is not conformally invariant. This takes us away from the  2d CFT framework which has been so useful to classify and compute in theories with boundaries. 

If we stay within the holographic regime of large $\beta \J$, then the situation simplifies. In this case the correct boundary conditions become $g(\tau_1,\tau_2) \sim \log \tau_{12}$ as $\tau_{12}\to 0$. This is an appropriate prescription as long as $\tau_{12}$ is small but still bigger than $1/\J$. Ignoring the correction when the two times become so close to each other is equivalent to considering ZZ-brane type boundary conditions on the 2d Liouville theory, as anticipated in \cite{MTV}.


\vspace{2mm}

\begin{center}
{\bf Acknowledgements}
\end{center}

\vspace{-2mm}

We thank A. Blommaert, N. Callebaut, S. Das, R. Dijkgraaf, Y. Gu, J. Kaplan, P. Nayak, X. L. Qi, J. Sonner, D. Stanford, M. Vielma, E. Witten and Z. Yang for useful discussions. H.T.L is supported by a Croucher Scholarship for Doctoral Study and a Centennial Fellowship from Princeton University.
T.M. acknowledges financial support from the Research Foundation-Flanders (FWO Vlaanderen). The research of H.V. is supported by NSF grant PHY-1620059.

\begin{appendix}
\def\WW{\mbox{\small W}}

\section{Rindler and Unruh modes}
\label{app:Unruh}
We collect some relevant formulas on the construction of Unruh modes and the Bogoliubov transformations, used in section \ref{sect:secondq}.
The Klein-Gordon inner product is defined for any Cauchy slice $\Sigma$ as:
\begin{equation}
\left(\phi_1,\phi_2\right) \equiv - i \int_{\Sigma} d\Sigma^\mu\left(\phi_1 \partial_\mu \phi_2^* - \phi_2^* \partial_\mu \phi_1\right).
\end{equation}
Normalizing all modes as
\begin{equation}
\left(\phi_i,\phi_j\right) = \delta_{ij}, \quad \left(\phi_i^*,\phi_j^*\right) = -\delta_{ij}, \quad \left(\phi_i,\phi_j^*\right) = \left(\phi_i^*,\phi_j\right) = 0,
\end{equation}
a (chiral component of a) massless field can be expanded in Rindler modes:
\begin{equation}
\phi(U) = \int_{0}^{+\infty} d\nu \left[a_{{\!}_{\rm R}}(\nu)  \psi_{{\!}_{\rm R \nu}} (U) + a^{\dagger}_{{\!}_{\rm R}}(\nu) \psi_{{\!}_{\rm R \nu}}^* (U) + a_{{\!}_{\rm L}}(\nu) \psi_{{\!}_{\rm L \nu}} (U) + a^{\dagger}_{{\!}_{\rm L}}(\nu) \psi_{{\!}_{\rm L \nu}}^* (U)\right],
\end{equation}
with
\bea
\label{2lrmodes}
 \psi_{{\!}_{\rm L\nu}} (U) = \frac{1}{\sqrt{4\pi\nu}}|U|^{- i\nu} \, \theta(-U), \qquad  \psi_{{\!}_{\rm R\nu}} (U) = \frac{1}{\sqrt{4\pi\nu}}U^{ i\nu} \, \theta(U),
\eea
or in Unruh modes:
\begin{equation}
\phi(U) = \int_{0}^{+\infty} d\nu \left[c^1(\nu) h^1_\nu(U) + c^{1\dagger}(\nu) h^{1*}_{\nu}(U) + c^2(\nu) h^2_\nu(U) + c^{2\dagger}(\nu) h^{2*}_\nu(U)\right],
\end{equation}                                                                                       
where $h^1_\nu$ and $h^2_\nu$ are two orthogonal Unruh modes that only contain positive Kruskal frequencies:
\begin{align}
\label{unruhmodes1}
h^1_\nu(U) &= \frac{i}{\sqrt{2\pi \nu}}\Gamma(1-i\nu)\left[e^{\frac{\pi\nu}{2}}\psi_{{\!}_{\rm R \nu}} (U) + e^{-\frac{\pi\nu}{2}}\psi_{{\!}_{\rm L \nu}}^* (U)\right], \\
\label{unruhmodes2}
h^2_\nu(U) &= -\frac{i}{\sqrt{2\pi \nu}}\Gamma(1+i\nu)\left[e^{\frac{\pi\nu}{2}}\psi_{{\!}_{\rm L \nu}} (U) + e^{-\frac{\pi\nu}{2}}\psi_{{\!}_{\rm R \nu}}^* (U)\right],
\end{align}
mainly supported in the $R$-, respectively $L$-wedge. 
These expansions are related by a Bogoliubov transformation:\footnote{One indeed quickly checks that 
\begin{align}
\left[c^1(\nu),c^{1\dagger}(\nu')\right] &= \left[c^2(\nu),c^{2\dagger}(\nu')\right] = \delta(\nu-\nu'), \quad \text{all others vanish}, \\
&\Updownarrow \nonumber \\
\left[a_{{\!}_{\rm R}}(\nu),a^{\dagger}_{{\!}_{\rm R}}(\nu')\right] &= \left[a_{{\!}_{\rm L}}(\nu),a^{\dagger}_{{\!}_{\rm L}}(\nu')\right] = \delta(\nu-\nu'), \quad \text{all others vanish}.
\end{align}}
\begin{align}
\label{bogoliubov}
a^{\dagger}_{{\!}_{\rm R}}(\nu) = \frac{-i\Gamma(1+i\nu)}{\sqrt{2\pi\nu}} \left( e^{\pi\nu/2}\,c^{1\dagger}(\nu) + e^{-\pi\nu/2}\,c^2(\nu)\right), \\
a_{{\!}_{\rm L}}(\nu) = \frac{-i\Gamma(1+i\nu)}{\sqrt{2\pi\nu}} \left( e^{-\pi\nu/2}\,c^{1\dagger}(\nu) + e^{\pi\nu/2}\,c^2(\nu)\right).
\end{align}
E.g. a Rindler mode can be expanded in Unruh and then subsequently into Kruskal (i.e. Minkowski) modes as\footnote{Useful formula:
\begin{align}
\int_{0}^{+\infty}dz e^{\pm ixz}z^{s-1} = e^{\pm i\frac{\pi}{2}s} x^{-s}\Gamma(s).
\end{align}}
\begin{align}
\psi_{{\!}_{\rm R \nu}} (U) &= \frac{1}{\sqrt{4\pi \nu}}U^{i\nu}\theta(U) \nonumber \\
&= -i\frac{e^{\pi\nu/2}\Gamma(1+i\nu)}{\sqrt{2\pi\nu}} h^1_\nu(U) + i\frac{e^{-\pi\nu/2}\Gamma(1+i\nu)}{\sqrt{2\pi\nu}} h^{2*}_\nu(U)\\
&= -i\frac{e^{\pi\nu/2}\Gamma(1+i\nu)}{2\pi\sqrt{\nu}} \int_{0}^{+\infty} dp\, p^{-i\nu-1/2} \psi_p(U) + i\frac{e^{-\pi\nu/2}\Gamma(1+i\nu)}{2\pi\sqrt{\nu}} \int_{-\infty}^{0} dp\, \left|p\right|^{-i\nu-1/2} \psi_p(U), \nonumber
\end{align}
with Kruskal mode $\psi_p(U) = \frac{e^{ipU}}{\sqrt{4\pi p}}$.\footnote{$U=X-T$ so this mode has indeed positive Kruskal frequency when $p>0$.} From this, we indeed identify the Kruskal content of the Unruh modes as $p^{-i\nu-1/2}\theta(p)$ and $\left|p\right|^{-i\nu-1/2}\theta(-p)$ as in \eqref{pmmodes}:
\begin{equation}
h^1_\nu(U) = \frac{1}{\sqrt{2\pi}} \int_{0}^{+\infty} dp\, p^{-i\nu-1/2} \psi_p(U), \qquad h^2_\nu(U) = \frac{1}{\sqrt{2\pi}} \int_{0}^{+\infty} dp\, p^{i\nu-1/2} \psi_p(U).
\end{equation}
Inspecting the modes \eqref{unruhmodes1},\eqref{unruhmodes2}, one can write the Unruh modes more economically by defining:
\begin{equation}
\mathbf{h}_\nu (U) \equiv \begin{cases} 
      h^1_{\nu}(U), \qquad \nu > 0 \\
      h^2_\nu(U) = h^1_{-\nu}(U),\, \nu < 0
   \end{cases}, \qquad 
\mathbf{c}(\nu) \equiv \begin{cases} 
      c^1(\nu), \qquad \nu > 0 \\
      c^2(\nu) = c^1(-\nu),\, \nu < 0
	 \end{cases}.
\end{equation}
Then the Unruh creation operator can be expanded into the Kruskal creation operators as Mellin transforms:
\begin{equation}
\label{Unruexpa}
\mathbf{c}^{\dagger}(\nu) = \frac{1}{\sqrt{2\pi}} \int_{0}^{+\infty} dp\, p^{-i\nu-1/2} a^\dagger_p, \qquad a^{\dagger}_p = \frac{1}{\sqrt{2\pi}} \int_{-\infty}^{+\infty} d\nu\, p^{i\nu-1/2} \mathbf{c}^\dagger(\nu),
\end{equation}
consistent with the Minkowski commutation relations:
\begin{equation}
\left[a_p,a^{\dagger}_{p'}\right] = \delta(p-p'), \quad \text{all others vanish}.
\end{equation}

The TFD state is, by definition, annihilated by all positive Kruskal frequency modes:
\begin{equation}
c^1(\nu)\left|\text{TFD}\right\rangle = c^2(\nu)\left|\text{TFD}\right\rangle = 0.
\end{equation}
To link the first and second quantized formalism, we should identify states through
\begin{equation}
a^{\dagger}_{{\!}_{\rm R}}(\nu) \left|0 \right\rangle = \frac{1}{\sqrt{2\pi\nu}} \left|\nu\right\rangle,
\end{equation}
leading to 
\begin{equation}
\label{tbog}
\sqrt{2\pi\nu} \, a^{\dagger}_{{\!}_{\rm R}}(\nu) \left|\text{TFD}\right\rangle 
\,\, \sim \,\, e^{\pi\nu/2}\Gamma(1+i\nu) \, c^{1\dagger}(\nu) \left|\text{TFD}\right\rangle,
\end{equation}
where the resulting states can finally be expanded into Kruskal eigenstates using \eqref{Unruexpa} as:
\begin{equation}
c^{1\dagger}(\nu) \left|\text{TFD}\right\rangle = \frac{1}{\sqrt{2\pi}} \int_{0}^{+\infty} dp\, p^{-i\nu-1/2} \left|p\right\rangle.
\end{equation}
At this point, one can make contact with the 't Hooft-Dray shockwave $\mathcal{S}$-matrix computation done in \eqref{smatf}.
The result \eqref{tbog} means there is an extra factor of $e^{\pi\nu/2}\Gamma(1+i\nu)$ when going from modes that are localized within either the $L$- or $R$-wedge, to a mode with positive Kruskal momentum.

An interesting example correlator to compute using the above formulas, is:
\begin{equation}
\left\langle \text{TFD}\right| a_{{\!}_{\rm L}}(-\nu_1)a^{\dagger}_{{\!}_{\rm R}}(\nu_2)\left|\text{TFD} \right\rangle  = \frac{1}{2\sinh(\pi \nu_1)} \delta(\nu_1-\nu_2),
\end{equation}
which is non-zero.

\section{Semiclassical limit of the R-matrix}\label{app:R}

\vspace{-2mm}

In this appendix we will collect some properties of the $R$-matrix that is involved in the out-of-time ordered correlators, needed to take the semiclassical limit of the Schwarzian theory. 
The R-matrix $ R_{k_sk_t}\! 
 \left[\, {}^{k_4}_{k_1} \,{}^{\Deltal_2}_{\Deltal_1}\right]$ is equal to a $6j$-symbol of $SU(1,1)$.
  It is given by the following  expression
 \bea
~~\mbox{\normalsize $R_{k_sk_t}\! 
 \left[\, {}^{k_4}_{k_2} \,{}^{\Deltal_2}_{\Deltal_1}\right]$}\is 
\sqrt{\frac{\Gamma(\Deltal_1+ik_2 \pm ik_s)\Gamma(\Deltal_2-ik_2\pm ik_t)\Gamma(\Deltal_1-i k_4 \pm ik_t) \Gamma(\Deltal_2+ik_4\pm ik_s)}{\Gamma(\Deltal_1-ik_2 \pm ik_s)\Gamma(\Deltal_2+ik_2\pm ik_t)\Gamma(\Deltal_1+i k_4 \pm ik_t) \Gamma(\Deltal_2-ik_4\pm ik_s)}}
 \nonumber \\[-2mm]\\[-2mm]
\! & \!\! & \!\hspace{-2.6cm}\times \int\limits_{-i\infty}^{i\infty}\!\! \frac{du}{2\pi i} \, \frac{\Gamma(u)\Gamma(u\! -\! 2ik_s)\Gamma(u\! +\!  i k_{2+4-s+t})\Gamma(u\! -\! i k_{s+t-2-4})\Gamma(\Deltal_1\!  + \! i k_{s-2}\! -\! u)\Gamma(\Deltal_2\! +\! i k_{s-4}\! -\! u)}{\Gamma(u\! +\! \Deltal_1\!  - \! i k_{s-2})\Gamma(u\! +\! \Deltal_2\! -\! ik_{s-4})} ,\nonumber
\label{appfinal}
\eea
where $k_{i+j}$  is shorthand for $k_i + k_j$. 
The integral over $u$ can be done by contour deformation to the right, yielding the Wilson function introduced by Groenevelt \cite{groenevelt}, which in turn can be expressed in terms of ${}_4 F_3$ hypergeometric functions. 

\begin{figure}[h]
\centering
\includegraphics[width=0.45\textwidth]{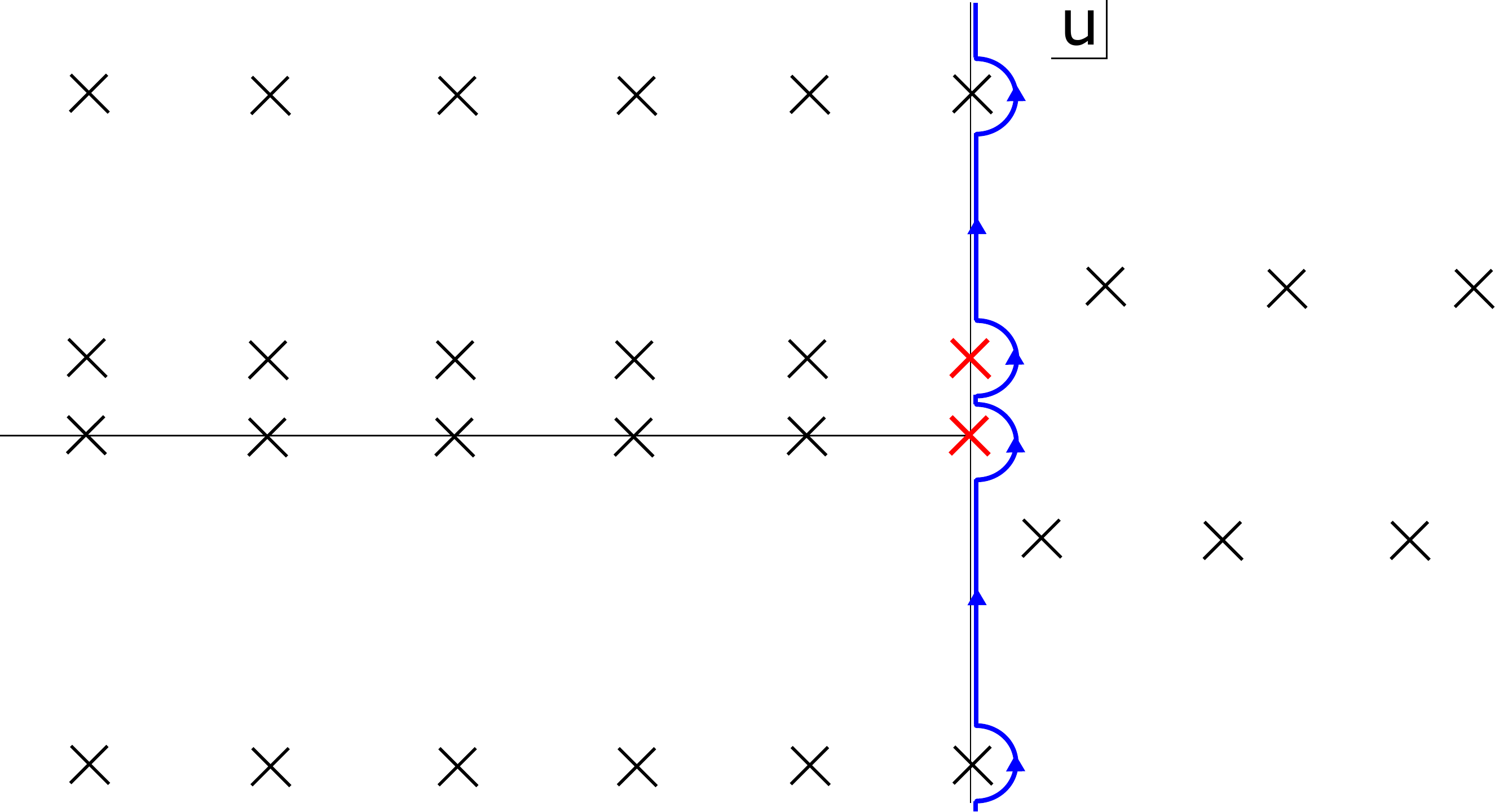}
\caption{Contour followed in the integral. Deforming to the right gives two ${}_4F_3$ functions or the Wilson function. Deforming to the left is more suitable to deduce that as $M\to \infty$ only 2 poles dominate (colored in red).}
\label{poles}
\end{figure}

To deduce the semi-classical regime however, it is more useful to deform the contour to the left instead. Consider the general integral
\begin{equation}
\int_{-i \infty}^{+i\infty} \frac{du}{2\pi i}~\frac{\Gamma(a_1+u)\Gamma(a_2+u)\Gamma(a_3+u)\Gamma(a_4+u)}{\Gamma(b_1+u)\Gamma(b_2+u)} ~\Gamma(A-u)\Gamma(B-u),
\end{equation}
for $a_1 \to i\infty$ and $a_3 \to -i\infty$ in the same way.\footnote{In detail:
\begin{align}
a_1 &= ik_{t-s+1+4}, \quad a_2 =-ik_{t+s-1-4}, \nonumber \\
a_3 &= -2ik_s, \quad a_4 = 0, \nonumber \\
b_1 &= j_1 - ik_{s-1}, \quad b_2 = j_2 - ik_{s-4}, \nonumber \\
A &= j_1 + ik_{s-1}, \quad B = j_2 + ik_{s-4}.
\end{align}} Deforming the contour to the left, we pick up poles from the first four $\Gamma$'s in the numerator. The poles are 4 series $u=-a_i-n$ starting at the imaginary axis and moving to the left (Figure \ref{poles}).
Two of these pole series are at $\Im(u) = \pm i \infty$. As $\Gamma(c \pm i \infty) \to 0$, these disappear in the limit, e.g. at $s=-a_1$, the residue equals
\begin{equation}
\frac{\Gamma(a_2-a_1)\Gamma(a_3-a_1)\Gamma(a_4-a_1)}{\Gamma(b_1-a_1)\Gamma(b_2-a_1)} ~\Gamma(A+a_1)\Gamma(B+a_1) \,\,\, \to \,\,\, 0,
\end{equation}
due to more $\Gamma$'s in the numerator than in the denominator: it is suppressed by 3 $\Gamma$'s, one of which is doubly suppressed. 

For the two remaining series of poles, all poles with $n\neq0$ are also subdominant due to $\Gamma(a_1-1) = \frac{\Gamma(a_1)}{a_1-1}$ and again more $\Gamma$'s in the numerator than in the denominator. E.g. at $u=-1$ the residue becomes:
\begin{equation}
\frac{\Gamma(a_1)\Gamma(a_2-1)\Gamma(a_3)}{(a_1-1)(a_3-1)\Gamma(b_1-1)\Gamma(b_2-1)} ~\Gamma(A+1)\Gamma(B+1) \,\,\, \to \,\,\, 0,
\end{equation}
which due to the additional $1/(a_1a_3)$ goes to zero much faster than the residue at $s=0$. \\
Two residues remain, at $u=0$ and at $u=-a_2$, each with half weight. These are suppressed by only 2 $\Gamma$'s, making them the dominant contribution. This proves the simplifying ansatz we made in \cite{MTV} to evaluate the $u$-integral in the semiclassical regime. 

The residues of both poles turn out to be related by
\begin{equation}
\text{Res}_{u=0} = \left.\text{Res}_{u=-a_2}\right|_{s \leftrightarrow -t, 1 \leftrightarrow -4 }.
\end{equation}
Focusing on the second pole, the relevant Gamma-functions in the amplitude are written as
\begin{equation}
\Gamma(\ell_1 + ik_1 - ik_s) \Gamma(\ell_1 + ik_4-ik_t) \Gamma(\ell_2+ik_1-ik_t) \Gamma(\ell_2+ik_4-ik_s)\Gamma(ik_s+ik_t-ik_1-ik_4).
\end{equation}
So all $\Gamma$'s just have a sign-flip in their dependence on all $k$'s compared to the $u=0$ pole. So upon defining the $\omega$'s with opposite sign as
\begin{align}
k_1^2 - k_s^2 = \omega_1, \quad k_4^2 - k_s^2 = \omega_4, \quad k_1^2-k_t^2 = \omega_3, \quad k_4^2 - k_t^2 = \omega_2, \quad k_s^2 = M,
\end{align}
one obtains in the end, comparing to the other pole, the time-reversed amplitude where every $t_i \to -t_i$.

\vspace{-2mm}

\section{Higher-point functions and multiple shockwaves}
\label{secthigh}

In this appendix we will show that the results from the previous sections directly generalize to arbitrary $2n$-point correlators. 
This serves both as an illustration of the general diagrammatic rules in section \ref{sectFR} in more complicated situations, and as a check on the semi-classical physics contained within the higher-point OTO correlators. Higher-order OTO correlation functions have been studied recently in \cite{Haehl:2017pak, Qi:2018rqm}.

We will prove that in the large $C$ regime, the Schwarzian correlation functions factorize into consecutive and independent 2-to-2 shockwave scattering processes. 
Moreover, the topology of the (real-time) shockwave graph is identical to the (Euclidean) Schwarzian diagram.

A  simple generalization from earlier sections is that all time-ordered correlation functions (those without any crossing lines in the graph) factorize into separate two point functions $G_{\ell}(\tau_{ij})$, generalizing this statement from time-ordered two- and four-point functions. 
This will also hold for pieces within OTO correlation functions, whose lines do not cross the remainder of the graph.

As examples of more complicated OTO correlation functions, we will analyze the following two diagrams and their semi-classical shockwave content:
\begin{align}
\begin{tikzpicture}[scale=0.65, baseline={([yshift=0cm]current bounding box.center)}]
\draw[thick]  (0,0) ellipse (1.6 and 1.6);
\draw[thick]  plot[smooth, tension=.7] coordinates {(1.6,0) (-1.05,1.2)};
\draw[thick]  plot[smooth, tension=.7] coordinates {(1.13,-1.13) (-1.6,0)};
\draw[thick]  plot[smooth, tension=.7] coordinates {(1.05,1.2) (-1.13,-1.13)};
\draw[fill,black] (-1.6,0) circle (0.1);
\draw[fill,black] (-1.05,1.2) circle (0.1);
\draw[fill,black] (1.6,0) circle (0.1);
\draw[fill,black] (1.05,1.2) circle (0.1);
\draw[fill,black] (-1.13,-1.13) circle (0.1);
\draw[fill,black] (1.13,-1.13) circle (0.1);
\end{tikzpicture}
\qquad\qquad
\begin{tikzpicture}[scale=0.65, baseline={([yshift=0cm]current bounding box.center)}]
\draw[thick]  (0,0) ellipse (1.6 and 1.6);
\draw[thick]  plot[smooth, tension=.7] coordinates {(1.48,0.61) (-0.61,-1.48)};
\draw[thick]  plot[smooth, tension=.7] coordinates {(0.61,1.48) (-1.48,-0.61)};
\draw[thick]  plot[smooth, tension=.7] coordinates {(1.48,-0.61) (-0.61,1.48)};
\draw[thick]  plot[smooth, tension=.7] coordinates {(0.61,-1.48) (-1.48,0.61)};
\draw[fill,black] (1.48,0.61) circle (0.1);
\draw[fill,black] (1.48,-0.61) circle (0.1);
\draw[fill,black] (0.61,1.48) circle (0.1);
\draw[fill,black] (0.61,-1.48) circle (0.1);
\draw[fill,black] (-1.48,0.61) circle (0.1);
\draw[fill,black] (-1.48,-0.61) circle (0.1);
\draw[fill,black] (-0.61,1.48) circle (0.1);
\draw[fill,black] (-0.61,-1.48) circle (0.1);
\end{tikzpicture}
\end{align}

\subsection{Example: double crossed diagram}
\begin{figure} 
\begin{center}
\includegraphics[width=0.25\textwidth]{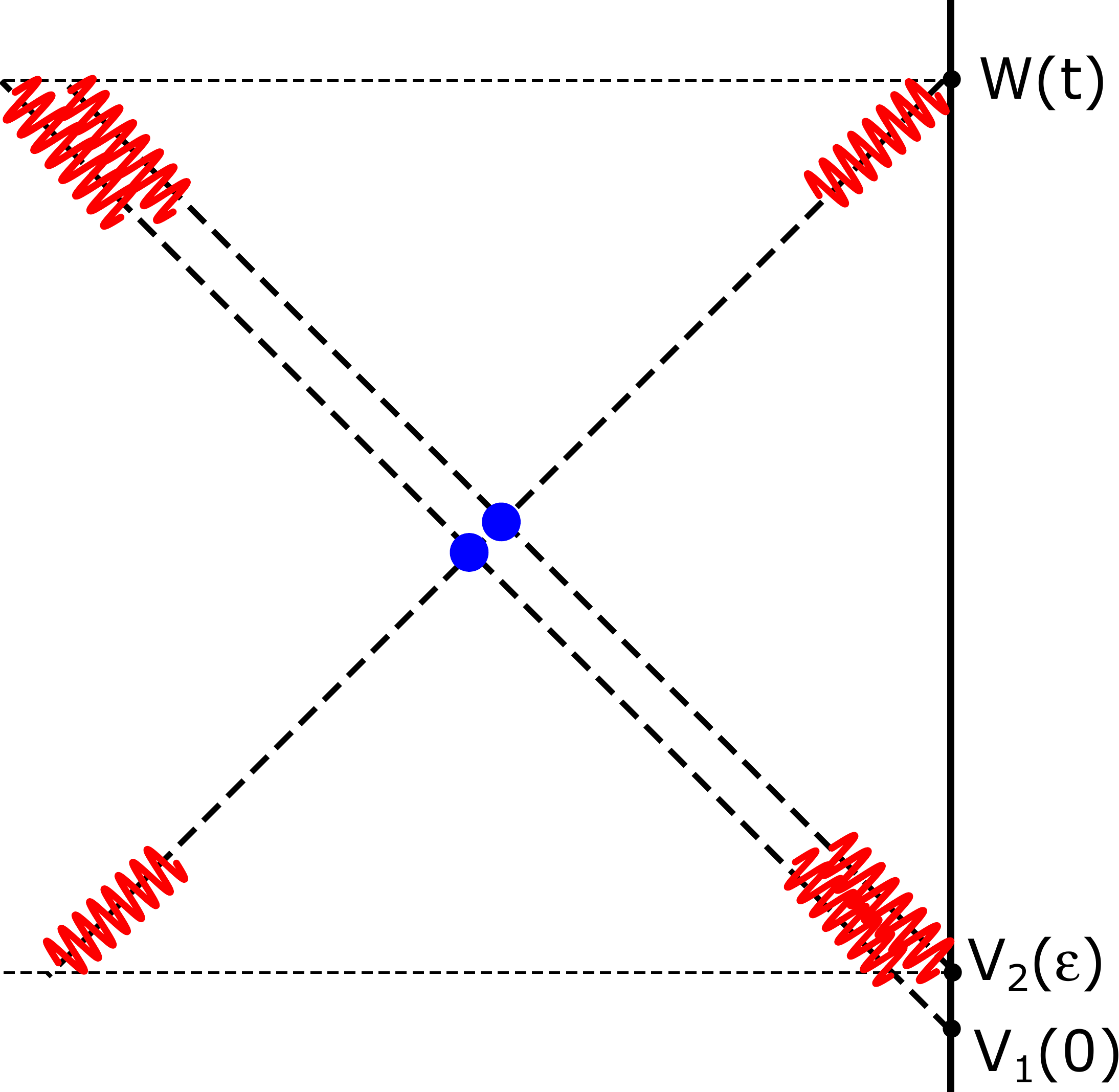}
\caption{Shockwave interaction between two infalling modes and one outgoing mode.}
\label{fig:shock}
\end{center}
\vspace{-4mm}
\end{figure}

As the first non-trivial example, we will analyze a double crossed diagram. This will correspond to a double shockwave process as represented by the semi-classical limit of the following OTO six-point function:
\begin{equation}
\left\langle W(t)V_1(0)V_2(\epsilon)W(t)V_2(\epsilon)V_1(0)\right\rangle,
\end{equation}
which can be represented as an $\mathcal{S}$-matrix overlap between in- and out-states: 
\begin{equation}
\left|\text{in} \right\rangle = W(t)V_2(\epsilon)V_1(0)\left|0\right\rangle, \quad \left|\text{out}\right\rangle = V^{\dagger}_2(\epsilon)V^{\dagger}_1(0)W^{\dagger}(t)\left|0\right\rangle.
\end{equation}
To interpret such a correlator in the Schwarzian theory, identical operators are connected into bilocal operators. This process is represented graphically in AdS$_2$ as shockwave scattering, as shown in figure \ref{fig:shock}.

The slightly more general situation of $n$ transverse lines (here $n=2$) is relevant when addressing wormhole traversability and was explored in \cite{MSY} in the large $C$ regime. Introducing again Kruskal momenta for each line, the relevant shockwave Dray-'t Hooft $\mathcal{S}$-matrix interaction is described by
\bea
\mathcal{S} \is \exp\Bigl(\kappa p_+\sum_{j=1}^n p_{j-}\Bigr).
\eea 
with $\kappa = \frac{i \beta}{4\pi C}$.
This interaction is described in Schwarzschild energies $\nu_i$ by the following expression (where $j_1,j_2$ are the two labels associated to all the crossing lines):
\begin{align}
\mathcal{S}(\nu_i) &= \int^{\infty}_0 \frac{dp_+}{p_+} p_+^{i(\nu_1-\nu_2)} \prod_{j}\int^{\infty}_0 \frac{dp_{j-}}{p_{j-}}p_{j-}^{-i(\nu_{j_1}-\nu_{j_2})}e^{\frac{i\beta}{4\pi C} p_{j-}p_+} \nonumber \\
&= 2\pi \delta\big(\nu_1-\nu_2+\sum_{j}(\nu_{j_1}-\nu_{j_2})\big) \, e^{\frac{\pi}{2}\sum_j (\nu_{j_1}-\nu_{j_2})}\kappa^{i\sum_j(\nu_{j_1}-\nu_{j_2})} \, \prod_{j}\Gamma(-i\nu_{j_1}+i\nu_{j_2}) .
\end{align}
We will reproduce this structure from the semi-classical limit of Schwarzian OTO correlators.

Within the Schwarzian theory, the double crossed diagram we want to analyze is
\begin{align}
\mathcal{A}_6^{\text{OTO}}(k_i,\ell_i,\tau_i) \, = \,\,\,
\begin{tikzpicture}[scale=0.7, baseline={([yshift=0cm]current bounding box.center)}]
\draw[thick]  (0,0) ellipse (1.6 and 1.6);
\draw[thick]  plot[smooth, tension=.7] coordinates {(1.6,0) (-1.05,1.2)};
\draw[thick]  plot[smooth, tension=.7] coordinates {(1.13,-1.13) (-1.6,0)};
\draw[thick]  plot[smooth, tension=.7] coordinates {(1.05,1.2) (-1.13,-1.13)};
\draw (-1.5,-1.5) node {\small $\tau_2$};
\draw (-1.5,1.5) node {\small $\tau_3$};
\draw (1.5,-1.5) node {\small $\tau_5$};
\draw (1.5,1.5) node {\small $\tau_1$};
\draw (2.2,0) node {\small $\tau_6$};
\draw (-2.2,0) node {\small $\tau_4$};
\draw[fill,black] (-1.6,0) circle (0.1);
\draw[fill,black] (-1.05,1.2) circle (0.1);
\draw[fill,black] (1.6,0) circle (0.1);
\draw[fill,black] (1.05,1.2) circle (0.1);
\draw[fill,black] (-1.13,-1.13) circle (0.1);
\draw[fill,black] (1.13,-1.13) circle (0.1);
\draw (0.45,1.15) node {\small $\ell_1$};
\draw (-0.75,0.7) node {\small $\ell_2$};
\draw (0.9,-.6) node {\small $\ell_3$};
\draw (0,2) node {\small \color{red}$k_2$};
\draw (0,-2) node {\small \color{red}$k_5$};
\draw (1.97,-.75) node {\small \color{red}$k_6$};
\draw (1.97,.75) node {\small \color{red}$k_1$};
\draw (-1.97,.75) node {\small \color{red}$k_{3}$};
\draw (-1.97,-.75) node {\small \color{red}$k_{4}$};
\end{tikzpicture}
\end{align}
This diagram corresponds to the amplitude (using the Feynman rules given in section \ref{sectFR}):
\begin{align}
&\mathcal{A}_6^{\text{OTO}}(k_i,\ell_i,t_i) = e^{-i\frac{t_{31}}{2C}k_2^2 - i\frac{t_{43}}{2C}k_{3}^2 - i\frac{t_{24}}{2C}k_{4}^2 - i\frac{t_{52}}{2C}k_5^2 -i\frac{t_{65}}{2C}k_6^2 + i\frac{t_{61}}{2C}k_1^2 - \frac{\beta}{2C}k_1^2} \\[2.5mm]
&\times \gamma_{\ell_1}(k_1,k_2)\gamma_{\ell_1}(k_4,k_5)\gamma_{\ell_2}(k_2,k_3)\gamma_{\ell_2}(k_1,k_6)\gamma_{\ell_3}(k_3,k_4)\gamma_{\ell_3}(k_5,k_6) \times R_{k_1k_{3}}\! 
 \left[\, {}^{k_6}_{k_2} \,{}^{\ell_2}_{\ell_1}\right] R_{k_6k_{4}}\! 
 \left[\, {}^{k_{5}}_{k_{3}} \,{}^{\ell_3}_{\ell_1}\right] \nonumber
\end{align}
Proceeding as in section \ref{sect5} by taking the residue of the $R$-matrix integrals, we can interpret the above expression as describing a six-point shockwave scattering diagram:
\begin{align}
\begin{tikzpicture}[scale=0.45, baseline={([yshift=0cm]current bounding box.center)}]
\draw[thick]  plot[smooth, tension=.7] coordinates {(3,-1) (-1,3)};
\draw[thick]  plot[smooth, tension=.7] coordinates {(1,-3) (-3,1)};
\draw[thick]  plot[smooth, tension=.7] coordinates {(3,3) (-3,-3)};
\draw[fill,blue] (1,1) circle (0.1);
\draw[fill,blue] (-1,-1) circle (0.1);
\draw[thick,<-] (2.0,-0.7) arc (-60:-30:2);
\draw[thick,<-] (0,-2.7) arc (-60:-30:2);
\draw[thick,->] (0,2.7) arc (120:150:2);
\draw[thick,->] (-2,0.7) arc (120:150:2);
\draw[thick,<-] (2.3,3) arc (60:30:2);
\draw[thick,->] (-2.3,-3) arc (240:210:2);
\draw (1,2.5) node {\small \color{red}$k_2$};
\draw (-1,-2.5) node {\small \color{red}$k_5$};
\draw (1.5,-1.5) node {\small \color{red}$k_6$};
\draw (2.5,1) node {\small \color{red}$k_1$};
\draw (-1.5,1.5) node {\small \color{red}$k_{3}$};
\draw (-2.5,-1) node {\small \color{red}$k_{4}$};
\end{tikzpicture}
\end{align}
The arrows depict the choice of sign of the $\omega$'s, so we define new redundant variables:
\begin{align}
k_2^2 - k_1^2 &= \omega_1, \quad k_6^2-k_1^2 = \omega_4, \quad k_5^2- k_6^2 = \omega_6, \nonumber \\
k_{4}^2-k_5^2 &= \omega_2, \quad k_{4}^2-k_{3}^2 = \omega_5, \quad k_3^2 - k_2^2 = \omega_3, \quad k_1^2 = M,
\end{align}
satisfying energy conservation $\omega_1+\omega_3+\omega_5 = \omega_2+\omega_4+\omega_6$, with as usual $M \gg \omega_i$ as $C \gg t_{ij}$. Defining again $\nu_i = \frac{\omega_i}{2\sqrt{M}}$, we find two final Gamma's associated to the two shockwave processes, written as
\begin{equation}
\Gamma\left(-i\nu_3+i\nu_4\right)\Gamma\left(-i\nu_5+i\nu_6\right),
\end{equation}
whereas the remainder gives precisely the required Schwarzschild wavefunctions \eqref{wvsch}:
\begin{align}
\sim \,\, &e^{i\nu_1 t_1}\frac{\Gamma(\ell_1 - i\nu_1)}{\sqrt{\Gamma(2\ell_1)}} \times  e^{- i\nu_2 t_2}\frac{\Gamma(\ell_1 + i\nu_2)}{\sqrt{\Gamma(2\ell_1)}} \times e^{i\nu_3 t_3}\frac{\Gamma(\ell_2 + i\nu_3)}{\sqrt{\Gamma(2\ell_2)}} \nonumber \\
&\times  e^{-i\nu_4 t_6}\frac{\Gamma(\ell_2 - i\nu_4)}{\sqrt{\Gamma(2\ell_2)}} \times e^{i\nu_5 t_4}\frac{\Gamma(\ell_3 + i\nu_5)}{\sqrt{\Gamma(2\ell_3)}} \times e^{-i\nu_6 t_5}\frac{\Gamma(\ell_3 - i\nu_6)}{\sqrt{\Gamma(2\ell_3)}}.
\end{align}

Finally, the $M$-integral, readily generalized to such an $n$-point OTO function, can be done as usual by saddle point methods, giving for any correlator the same saddle $M_0 = 2\pi^2 C/\beta^2$ as found before.

As alluded to already several times, the structure of this computation is immediately generalized to arbitrary $n$-point OTO crossed diagrams of this specific graph topology. 

\subsection{Example: quadruple crossed diagram}
A second non-trivial example of these rules is the quadruple crossed diagram depicted below:
\bea
\begin{tikzpicture}[scale=0.75, baseline={([yshift=0cm]current bounding box.center)}]
\draw[thick]  (0,0) ellipse (1.6 and 1.6);
\draw[thick]  plot[smooth, tension=.7] coordinates {(1.48,0.61) (-0.61,-1.48)};
\draw[thick]  plot[smooth, tension=.7] coordinates {(0.61,1.48) (-1.48,-0.61)};
\draw[thick]  plot[smooth, tension=.7] coordinates {(1.48,-0.61) (-0.61,1.48)};
\draw[thick]  plot[smooth, tension=.7] coordinates {(0.61,-1.48) (-1.48,0.61)};
\draw[fill,black] (1.48,0.61) circle (0.1);
\draw[fill,black] (1.48,-0.61) circle (0.1);
\draw[fill,black] (0.61,1.48) circle (0.1);
\draw[fill,black] (0.61,-1.48) circle (0.1);
\draw[fill,black] (-1.48,0.61) circle (0.1);
\draw[fill,black] (-1.48,-0.61) circle (0.1);
\draw[fill,black] (-0.61,1.48) circle (0.1);
\draw[fill,black] (-0.61,-1.48) circle (0.1);
\draw (-0.65,0.65) node {\small $\ell_1$};
\draw (0.65,-0.65) node {\small $\ell_2$};
\draw (0.65,0.65) node {\small $\ell_3$};
\draw (-0.65,-0.65) node {\small $\ell_4$};
\draw (2,0) node {\small \color{red}$k_1$};
\draw (1.42,1.42) node {\small \color{red}$k_2$};
\draw (0,2) node {\small \color{red}$k_3$};
\draw (-1.42,1.42) node {\small \color{red}$k_4$};
\draw (-2,0) node {\small \color{red}$k_5$};
\draw (-1.42,-1.42) node {\small \color{red}$k_6$};
\draw (0,-2) node {\small \color{red}$k_7$};
\draw (1.42,-1.42) node {\small \color{red}$k_8$};
\draw (0,0) node {\small \color{red}$k_9$};
\end{tikzpicture}
\raisebox{-3pt}{$\  = \ \ R_{k_2k_{4}}\! 
 \left[\, {}^{k_9}_{k_3} \,{}^{\ell_3}_{\ell_1}\right] R_{k_1k_{9}}\! 
 \left[\, {}^{k_8}_{k_{2}} \,{}^{\ell_3}_{\ell_2}\right] R_{k_8k_{6}}\! 
 \left[\, {}^{k_7}_{k_{9}} \,{}^{\ell_4}_{\ell_2}\right] R_{k_9k_{5}}\! 
 \left[\, {}^{k_6}_{k_{4}} \,{}^{\ell_4}_{\ell_1}\right]$}~~~ \\
\times \gamma_{\ell_1}(k_2,k_3)\gamma_{\ell_1}(k_5,k_6)\gamma_{\ell_2}(k_1,k_2)\gamma_{\ell_2}(k_6,k_7)\gamma_{\ell_3}(k_3,k_4)\gamma_{\ell_3}(k_1,k_8)\gamma_{\ell_4}(k_4,k_5)\gamma_{\ell_4}(k_7,k_8) \nonumber
\eea
It contains four $R$-matrices as it is obtained by four swaps of lines from the striped diagram with four rungs, as in \eqref{6ptdiagr}. The full amplitude is thus a nine-fold integral over $d\mu(k_i), i=1..9$ weighted with these four $R$-matrices, eight vertex functions, and with eight propagators (not written here). Note that the $k_9$-integral does not contain a propagator piece, and appears only within the four $R$-matrices. \\
The diagram is associated to the real-time OTO correlator:
\begin{equation}
\left\langle W_2(t+\epsilon)W_1(t)V_1(0)V_2(\epsilon)W_1(t)W_2(t+\epsilon)V_2(\epsilon)V_1(0)\right\rangle,
\end{equation}
which has an in-out $\mathcal{S}$-matrix interpretation of four 2-to-2 shockwave interactions, and indeed requires four swaps to bring the $W$-operators in-time ordering and untangle the correlator. \\
Understanding the semi-classical regime of this correlator requires all of the techniques as presented above for other (easier) diagrams. The additional novelty in this case is the $k_9$-integral over the four $R$-matrices, which in the semi-classical regime boils down to Barnes' first lemma and results in the semiclassical $\mathcal{S}$-matrix analogous to \eqref{smatf}, in agreement with the eikonal shockwave computation:
\begin{align}
\mathcal{S}(\nu_i) &\sim \, 2\pi \delta\bigl(-\nu_1+\nu_2-\nu_3+\nu_4-\nu_5+\nu_6-\nu_7+\nu_8\bigr) e^{-\frac{\pi}{2}(\nu_1-\nu_2+\nu_3-\nu_4)} \kappa^{-i(\nu_1-\nu_2+\nu_3-\nu_4)} \nonumber \\[2mm]
&\quad\quad \times \frac{\Gamma(i\nu_1-i\nu_2)\Gamma(i\nu_3-i\nu_4)\Gamma(-i\nu_5+i\nu_6)\Gamma(-i\nu_7+i\nu_8)}{\Gamma(i\nu_1-i\nu_2+i\nu_3-i\nu_4)} \nonumber \\[2mm]
&= \, \int_{0}^{+\infty}\frac{dp_{1+}dp_{2+}}{p_{1+}p_{2+}}p_{1+}^{i(\nu_1-\nu_2)}p_{2+}^{i(\nu_3-\nu_4)}\int_{0}^{+\infty}\frac{dp_{1-}dp_{2-}}{p_{1-}p_{2-}}p_{1-}^{-i(\nu_5-\nu_6)}p_{2-}^{-i(\nu_7-\nu_8)} \nonumber \\[2mm]
&\quad\quad \times e^{i \kappa (q_1p_1+q_1p_2+q_2p_1+q_2p_2)}.
\end{align}

\medskip

\section{Heavy two-point function from the Schwarzian saddle}

\vspace{-2mm}

\begin{figure}[t]
\begin{minipage}{0.45\textwidth}
\centering
\includegraphics[width=0.9\textwidth]{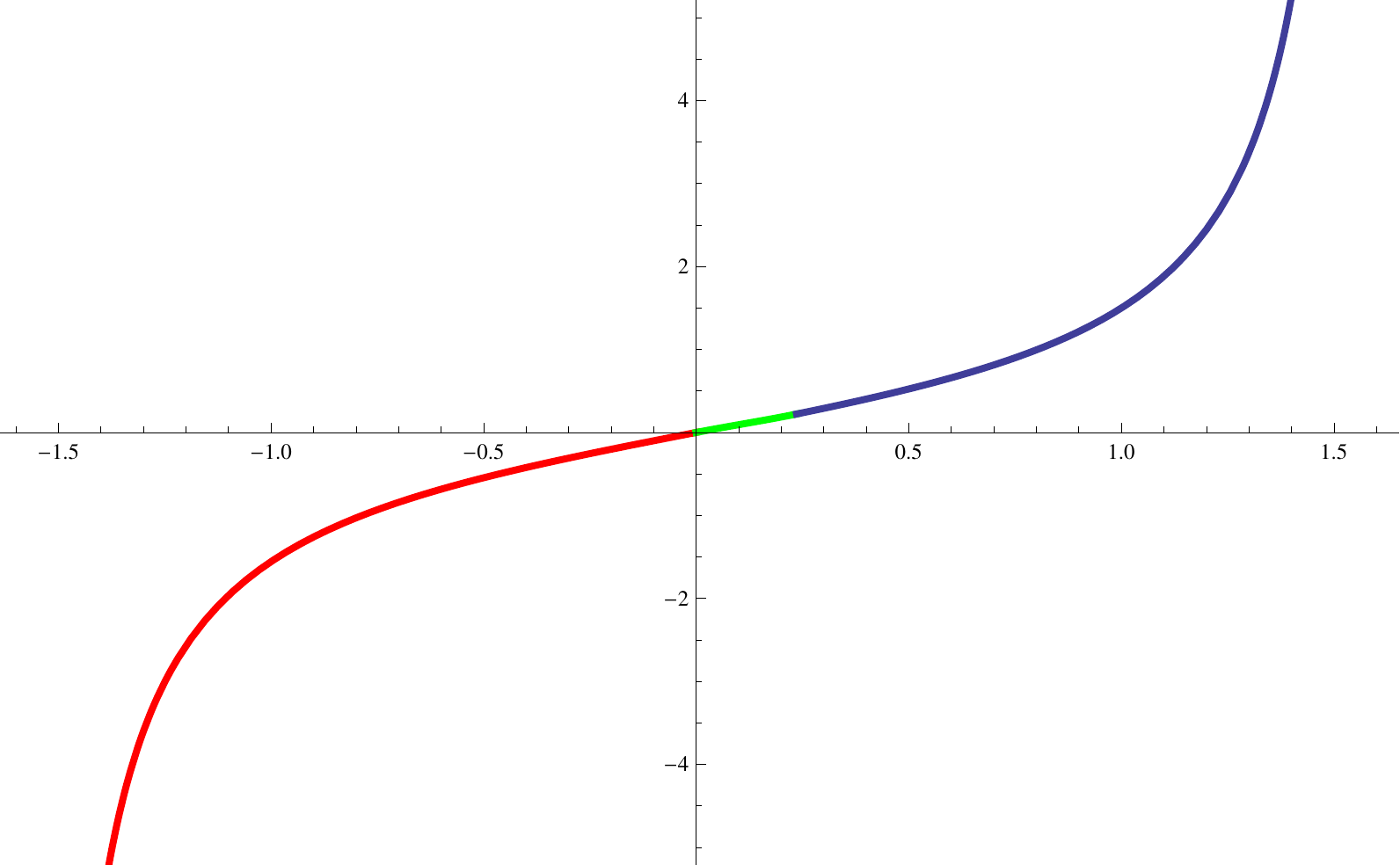}
\caption{Full solution with non-zero temperature, with $\tau_1=0$, $\ell = 2C$, $k_2 = 1$, $k_1 = 3$, with saddle values $\tau_2 \approx 0.236$, $\tau_f \approx 1.588$ and $\beta\approx 3.159$.}
\label{ClassicalHeavyLimitT}
\end{minipage}
\hspace{0.075\textwidth}
\begin{minipage}{0.45\textwidth}
\centering
\includegraphics[width=0.9\textwidth]{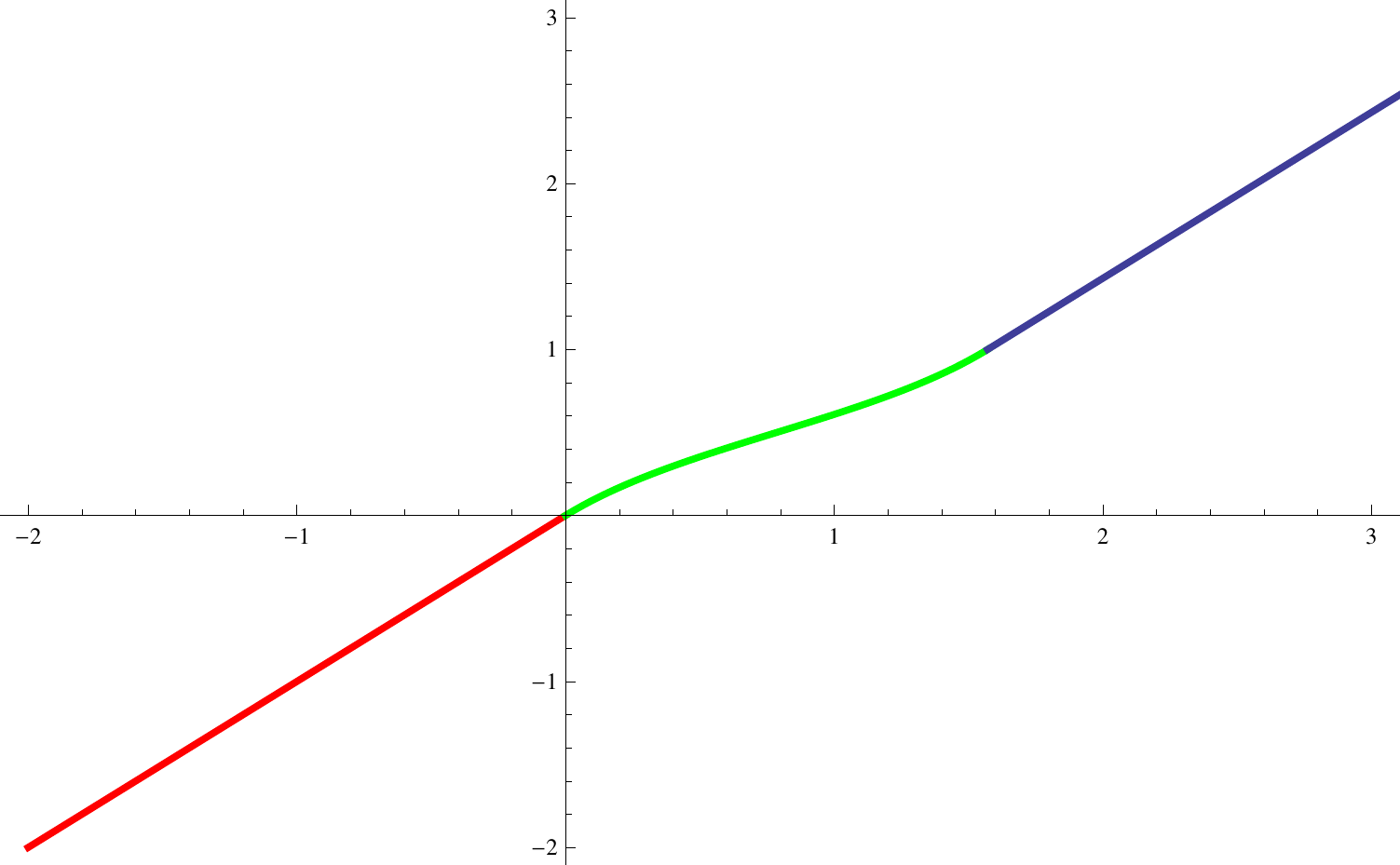}
\caption{Full solution at zero temperature with $t_1=0$, $t_2 = \frac{\pi}{2}$, $\ell = 2C$, $\sqrt{\frac{E}{2C}} = 1$.}
\label{ClassicalHeavyLimit}
\end{minipage}
\end{figure}

\label{sectheavy}
We complement the semiclassical limit of heavy two-point correlators of section \ref{sectell}, by comparing these results to an explicit solution of the Schwarzian equations of motion following from \eqref{semipath}. In the semi-classical limit, the path integral \eqref{semipath} is dominated by the classical solution to the action.
Using that 
$
\delta \left\{F,\tau\right\} 
= -\frac{\left\{F,\tau\right\}'}{F'} \delta F$, the equation of motion of $f$ gives
\begin{equation}
\label{eqmot}
C\frac{\left\{F,\tau\right\}'}{F'} + \ell \left[\frac{\delta'(\tau-\tau_1)}{F'_1} + \frac{\delta'(\tau-\tau_2)}{F'_2} + \frac{2}{F_1-F_2}\left[\delta(\tau-\tau_1)-\delta(\tau-\tau_2)\right]\right] = 0.
\end{equation}
After integrating the equation once, we find that the energy $E = C\{ F,\tau\}$ is piecewise constant 
\begin{alignat}{2}
\label{energyeq}
C\{ F,\tau\} &= E_2, \quad &&\tau<\tau_1, \nonumber \\[1mm]
C\{ F,\tau\} &= E_2 + 2\ell\frac{F'_1}{F_2-F_1} + \ell \frac{F''_1}{F'_1} \; = \; E_1, \quad &&\tau_1<\tau<\tau_2, \\
C\{ F,\tau\}  &= E_2 + 2\ell \frac{F'_1-F'_2}{F_2-F_1} + \ell \frac{F''_1}{F'_1} + \ell \frac{F''_2}{F'_2}\; = \; E_2, \qquad &&\tau_2 < \tau. \nonumber
\end{alignat}
The quantities written here are averaged, e.g. $F''_1 \equiv \frac{F''^{<}_1 + F''^{>}_1}{2}$. The boundary conditions at $\tau_1$ and $\tau_2$ are read off from \eqref{eqmot}. 
Using the notation $T(u) = -\frac{\ell}{F'_1}\delta'(u-\tau_1) - \frac{2}{F_1-F_2}\delta(u-\tau_1)$, we can integrate \eqref{eqmot} into:
\beq
\frac{F''}{F'} = \int^t \!\! du \, (F(\tau)-F(u)) T(u), \qquad \qquad
F' = \int^t\!\! du \, \frac{(F(\tau)-F(u))^2}{2} T(u).
\eeq
A $\delta$-function insertion in $T(u)$ requires a jump for $F'''$, whereas $\delta'$-insertions require a jump already for $F''$. In this case, the gluing conditions at $\tau=\tau_1, \tau_2$ are
\beq
F, F' \text{ continuous}, \qquad \quad
\Delta F'' = -\frac{\ell}{C}F'.
\eeq
We set $\tau_1=0$ without loss of generality. For the combined solution, we make the following Ansatz
\begin{alignat}{2}
F(\tau) & = \, \frac {2C} {k_2}\tan\Bigl(\frac{k_2\spc \tau} {2C}\Bigr),\qquad  \quad &&\tau<0, \nonumber \\[-4mm]
& & & \qquad \qquad \qquad \qquad 
k_2 \, = \, \sqrt{2E_2 C},\nonumber \\[-2mm]
&= \, \frac{\tan\Bigl(\frac{k_1\spc \tau} {2C}\Bigr)}{\frac{\ell}{2C}\tan\Bigl(\frac{k_1\spc \tau} {2C}\Bigr)+\frac{k_1}{2C}},\qquad \quad &&0 <\tau<\tau_2, \label{zerotsolution}\\[-2mm]
&  & & \qquad \qquad \qquad \qquad 
k_1 \, = \, \sqrt{2E_1 C}, \nonumber \\[-3.3mm]
&= \, \frac{a \tan\left(\frac{k_2} {2C}(\tau-\tau_2)\right) + b \, \frac{k_2}{2C}}{ c \tan\left(\frac{k_2} {2C}(\tau-\tau_2)\right) + d\, \frac{k_2} {2C}} ,\qquad \quad &&\tau_2 < \tau.\nonumber
\end{alignat}
Setting $d=1$ and imposing the continuity conditions gives that
\begin{equation}
 b = F_2, \quad c = -\frac{1}{2}\frac{F''_2}{F_2} = -\frac{1}{2}\frac{F''^{<}_2}{F_2'} + \frac{\ell}{2C}, \quad a = F'_2 + F_2 c.
\end{equation}
From this point onwards, it is a straightforward calculation to show that the continuity conditions and the requirement that the distance between the asymptotes of $f(\tau)$ are fixed by the inverse temperature $\beta$ are precisely equivalent to the conditions \eqref{finitetsaddle}. An example of a classical solution is depicted in Figure~\ref{ClassicalHeavyLimitT}.

The finite temperature on-shell action is
\begin{align}
S_{0} &= -\frac{\tau k_1^{2}}{2C} - \frac{(\beta-\tau) k_2^{2}}{2C} - \ell \ln \frac{F'_2}{F_2^2},
\end{align}
where
\begin{equation}
\frac{F'_2}{F_2^2} = \frac{1}{(4C\ell)^2}((k_1+k_2)^2+\ell^2)((k_1-k_2)^2+\ell^2)
\end{equation}
agreeing with \eqref{sadintT}. 

These equations contain interesting bulk gravitational physics that is most easily seen in the zero-temperature limit (Figure \ref{ClassicalHeavyLimit}). The solution $f(\tau)$ is linear before and after the bilocal insertion, and is thermal in between. The net effect of the bilocal operator on the solution is a Shapiro time delay, corresponding to a mass $\sim \ell$ being injected and extracted in an otherwise vacuum space. The clock $f(\tau)$ is then delayed as it passes through this massive region. If $\ell < 0$, a time advance would be found.

\end{appendix}


\begin{thebibliography}{99}

	\bibitem{Dray:1984ha}
  T.~Dray and G.~'t Hooft,
  ``The Gravitational Shock Wave of a Massless Particle,''
  Nucl.\ Phys.\ B {\bf 253} (1985) 173.
  
	
	\bibitem{Kiem:1995iy}
  Y.~Kiem, H.~L.~Verlinde and E.~P.~Verlinde,
  ``Black hole horizons and complementarity,''
  Phys.\ Rev.\ D {\bf 52} (1995) 7053;
  K.~Schoutens, H.~L.~Verlinde and E.~P.~Verlinde,
  ``Quantum black hole evaporation,''
  Phys.\ Rev.\ D {\bf 48} (1993) 2670
  

\bibitem{SS} S.~H.~Shenker and D.~Stanford, ``Black holes and the butterfly effect,'' JHEP {\bf 1403}, 067 (2014)
  \href{http://arxiv.org/abs/1306.0622}{{\ttfamily arXiv:1306.0622 [hep-th]}}; ``Multiple Shocks,'' JHEP {\bf 1412}, 046 (2014) \href{http://arxiv.org/abs/1312.3296}{{\ttfamily arXiv:1312.3296 [hep-th]}}.
	
\bibitem{JMV} 
  S.~Jackson, L.~McGough and H.~Verlinde,
  ``Conformal Bootstrap, Universality and Gravitational Scattering,''
  Nucl.\ Phys.\ B {\bf 901}, 382 (2015)
  \href{https://arxiv.org/abs/1412.5205}{{\ttfamily [arXiv:1412.5205[hep-th]]}}. 

	
	     \bibitem{Shenker:2014cwa} 
  S.~H.~Shenker and D.~Stanford,
  ``Stringy effects in scrambling,''
  JHEP {\bf 1505}, 132 (2015)
  \href{https://arxiv.org/abs/1412.6087}{{\ttfamily [arXiv:1412.6087 [hep-th]]}}.
	
\bibitem{KitaevTalks}
A. Kitaev, Talk given at the Fundamental Physics Prize Symposium, \href{https://www.youtube.com/watch?v=OQ9qN8j7EZI}{Nov. 10, 2014}; A. Kitaev, KITP
seminar, \href{http://online.kitp.ucsb.edu/online/joint98/kitaev/}{Feb. 12, 2015}; ``A simple model of quantum holography,'' talks at KITP,  \href{http://online.kitp.ucsb.edu/online/entangled15/kitaev/}{April 7, 2015} and \href{http://online.kitp.ucsb.edu/online/entangled15/kitaev2/}{May 27, 2015}.

\bibitem{KS} 
A.~Kitaev and S.~J.~Suh,
  ``The soft mode in the Sachdev-Ye-Kitaev model and its gravity dual,''
  JHEP {\bf 1805}, 183 (2018)
 \href{http://arxiv.org/abs/1711.08467}{{\ttfamily arXiv:1711.08467 [hep-th]}}.

  
  \bibitem{MSS}J.~Maldacena, S.~H.~Shenker and D.~Stanford, ``A bound on chaos,''  \href{http://arxiv.org/abs/1503.01409}{{\ttfamily arXiv:1503.01409 [hep-th]}}
      
\bibitem{Sachdev:1992fk}
S.~Sachdev and J.-w. Ye, ``{Gapless spin fluid ground state in a random,
  quantum Heisenberg magnet},''
 Phys. Rev. Lett.
  {\bfseries 70} (1993) 3339,
\href{http://arxiv.org/abs/cond-mat/9212030}{{\ttfamily arXiv:cond-mat/9212030
  [cond-mat]}}.
  
\bibitem{Polchinski:2016xgd} 
  J.~Polchinski and V.~Rosenhaus,
  ``The Spectrum in the Sachdev-Ye-Kitaev Model,''
  JHEP {\bf 1604}, 001 (2016)
 \href{http://arxiv.org/abs/1601.06768}{{\ttfamily arXiv:1601.06768 [hep-th]}}.


  
\bibitem{Maldacena:2016hyu} 
  J.~Maldacena and D.~Stanford,
  ``Remarks on the Sachdev-Ye-Kitaev model,''
Phys.\ Rev.\ D {\bf 94}, no. 10, 106002 (2016)
\href{http://arxiv.org/abs/1604.07818}{{\ttfamily arXiv:1604.07818 [hep-th]}}. 

\bibitem{Jevicki:2016bwu} 
  A.~Jevicki, K.~Suzuki and J.~Yoon,
  ``Bi-Local Holography in the SYK Model,''
 JHEP {\bf 1607}, 007 (2016)
  \href{http://arxiv.org/abs/1603.06246}{{\ttfamily arXiv:1603.06246 [hep-th]}}; 
  A.~Jevicki and K.~Suzuki,
  ``Bi-Local Holography in the SYK Model: Perturbations,''
  JHEP {\bf 1611}, 046 (2016)
  \href{http://arxiv.org/abs/1608.07567}{{\ttfamily arXiv:1608.07567 [hep-th]}}.



\bibitem{Cotler:2016fpe} 
  J.~S.~Cotler {\it et al.},
  ``Black Holes and Random Matrices,''
  JHEP {\bf 1705}, 118 (2017)
   \href{http://arxiv.org/abs/1611.04650}{{\ttfamily arXiv:1611.04650 [hep-th]}}.


\bibitem{Almheiri:2014cka} 
  A.~Almheiri and J.~Polchinski,
  ``Models of AdS$_{2}$ backreaction and holography,''
  JHEP {\bf 1511}, 014 (2015)
  \href{http://arxiv.org/abs/1402.6334}{{\ttfamily arXiv:1402.6334 [hep-th]}};  R.~Jackiw,
  ``Lower Dimensional Gravity,''
  Nucl.\ Phys.\ B {\bf 252}, 343 (1985);  C.~Teitelboim,
  ``Gravitation and Hamiltonian Structure in Two Space-Time Dimensions,''
  Phys.\ Lett.\  {\bf 126B}, 41 (1983).



  \bibitem{Jensen:2016pah}
  K.~Jensen,
  ``Chaos and hydrodynamics near AdS$_2$,''
  \href{http://arxiv.org/abs/1605.06098}{{\ttfamily arXiv:1605.06098 [hep-th].}}

\bibitem{Maldacena:2016upp} 
  J.~Maldacena, D.~Stanford and Z.~Yang,
  ``Conformal symmetry and its breaking in two dimensional Nearly Anti-de-Sitter space,''
  PTEP {\bf 2016}, no. 12, 12C104 (2016)
 \href{http://arxiv.org/abs/1606.01857}{{\ttfamily arXiv:1606.01857 [hep-th]}}.

\bibitem{Engelsoy:2016xyb} 
  J.~Engelsoy, T.~G.~Mertens and H.~Verlinde,
  ``An investigation of AdS$_{2}$ backreaction and holography,''
 JHEP {\bf 1607}, 139 (2016)
 \href{http://arxiv.org/abs/1606.03438}{{\ttfamily arXiv:1606.03438 [hep-th]}}.
 
\bibitem{Cvetic:2016eiv} 
  M.~Cvetic and I.~Papadimitriou,
  ``AdS$_{2}$ holographic dictionary,''
  JHEP {\bf 1612}, 008 (2016)
  Erratum: [JHEP {\bf 1701}, 120 (2017)]
   \href{http://arxiv.org/abs/1608.07018}{{\ttfamily arXiv:1608.07018 [hep-th]}}.
  
		\bibitem{Nayak:2018qej}
  P.~Nayak, A.~Shukla, R.~M.~Soni, S.~P.~Trivedi and V.~Vishal,
  ``On the Dynamics of Near-Extremal Black Holes,''
  \href{https://arxiv.org/abs/1802.09547}{{\ttfamily [arXiv:1802.09547[hep-th]]}}. 
 

\bibitem{MTV} 
  T.~G.~Mertens, G.~J.~Turiaci and H.~L.~Verlinde,
  ``Solving the Schwarzian via the Conformal Bootstrap,''
  JHEP {\bf 1708}, 136 (2017)
  \href{http://arxiv.org/abs/1705.08408}{{\ttfamily [arXiv:1705.08408 [hep-th]]}}.
	
	 \bibitem{Turiaci:2016cvo} 
  G.~Turiaci and H.~Verlinde,
  ``On CFT and Quantum Chaos,''
  JHEP {\bf 1612}, 110 (2016)
      \href{http://arxiv.org/abs/1603.03020}{{\ttfamily arXiv:1603.03020  [hep-th]}}.
	
\bibitem{Turiaci:2017zwd} 
  G.~Turiaci and H.~Verlinde,
  ``Towards a 2d QFT Analog of the SYK Model,''
     \href{http://arxiv.org/abs/1701.00528}{{\ttfamily arXiv:1701.00528  [hep-th]}}. 

\bibitem{altland} 
   D.~Bagrets, A.~Altland and A.~Kamenev, ``Sachdev-Ye-Kitaev model as Liouville quantum mechanics,''
  Nucl.\ Phys.\ B {\bf 911}, 191 (2016)
   \href{http://arxiv.org/abs/1607.00694}{{\ttfamily arXiv:1607.00694 [cond-mat.str-el]}};
  ``Power-law out of time order correlation functions in the SYK model,''
  \href{http://arxiv.org/abs/1702.08902}{{\ttfamily arXiv:1702.08902 [cond-mat.str-el]}}.
	
	\bibitem{StWi}
  D.~Stanford and E.~Witten,
  ``Fermionic Localization of the Schwarzian Theory,''
  JHEP {\bf 1710}, 008 (2017)
  \href{http://arxiv.org/abs/1703.04612}{{\ttfamily arXiv:1703.04612 [hep-th]}}.
	
 \bibitem{Zhenbin}
  Z.~Yang,
  ``The Quantum Gravity Dynamics of Near Extremal Black Holes,''
  arXiv:1809.08647 [hep-th].

\bibitem{GJW}
  P.~Gao, D.~L.~Jafferis and A.~Wall,
  ``Traversable Wormholes via a Double Trace Deformation,''
  \href{https://arxiv.org/abs/1608.05687}{{\ttfamily arXiv:1608.05687 [hep-th]}}.

\bibitem{MSY}
  J.~Maldacena, D.~Stanford and Z.~Yang,
  ``Diving into traversable wormholes,''
  Fortsch.\ Phys.\  {\bf 65} (2017) no.5,  1700034
  \href{https://arxiv.org/abs/1704.05333}{{\ttfamily[arXiv:1704.05333 [hep-th]]}}.
	
		\bibitem{Hooft:2015jea}
  G.~'t Hooft,
  ``Diagonalizing the Black Hole Information Retrieval Process,''
  \href{https://arxiv.org/abs/1509.01695}{{\ttfamily arXiv:1509.01695 [gr-qc]}}.
	
	\bibitem{Hooft:2016itl}
  G.~'t Hooft,
  ``Black hole unitarity and antipodal entanglement,''
  Found.\ Phys.\  {\bf 46} (2016) no.9,  1185
  \href{https://arxiv.org/abs/1601.03447}{{\ttfamily [arXiv:1601.03447 [gr-qc]]}}.
	
	\bibitem{Unruh:1976db}
  W.~G.~Unruh,
  ``Notes on black hole evaporation,''
  Phys.\ Rev.\ D {\bf 14} (1976) 870.
  
  \bibitem{thomas} 
  T.~G.~Mertens,
  ``The Schwarzian Theory - Origins,''
  JHEP {\bf 1805} (2018) 036
 \href{https://arxiv.org/abs/1801.09605}{{\ttfamily [arXiv:1801.09605[hep-th]]}}. 

\bibitem{Zamolodchikov:2001ah}
  A.~B.~Zamolodchikov and A.~B.~Zamolodchikov,
  ``Liouville field theory on a pseudosphere,''
  \href{https://arxiv.org/abs/hep-th/0101152}{{\ttfamily [arXiv:0101152[hep-th]]}}.
		
			\bibitem{Gervais:1981gs}
  J.~L.~Gervais and A.~Neveu,
  ``The Dual String Spectrum in Polyakov's Quantization. 1.,''
  Nucl.\ Phys.\ B {\bf 199} (1982) 59.
  	
		\bibitem{Gervais:1982nw}
  J.~L.~Gervais and A.~Neveu,
  ``Dual String Spectrum in Polyakov's Quantization. 2. Mode Separation,''
  Nucl.\ Phys.\ B {\bf 209} (1982) 125.
  
	
	\bibitem{Gervais:1982yf}
  J.~L.~Gervais and A.~Neveu,
  ``New Quantum Solution of Liouville Field Theory,''
  Phys.\ Lett.\  {\bf 123B} (1983) 86.
  	
	\bibitem{Gervais:1983am}
  J.~L.~Gervais and A.~Neveu,
  ``New Quantum Treatment of Liouville Field Theory,''
  Nucl.\ Phys.\ B {\bf 224} (1983) 329.
  
	
			\bibitem{Dorn:2006ys}
  H.~Dorn and G.~Jorjadze,
  ``Boundary Liouville theory: Hamiltonian description and quantization,''
  SIGMA {\bf 3} (2007) 012
  \href{https://arxiv.org/abs/hep-th/0610197}{{\ttfamily [arXiv:0610197[hep-th]]}}.
	
	\bibitem{Dorn:2008sw}
  H.~Dorn and G.~Jorjadze,
  ``Operator Approach to Boundary Liouville Theory,''
  Annals Phys.\  {\bf 323} (2008) 2799
    \href{https://arxiv.org/abs/0801.3206}{{\ttfamily [arXiv:0801.3206[hep-th]]}}.

	
	  \bibitem{PT}
B.~Ponsot and J.~Teschner,
  ``Liouville bootstrap via harmonic analysis on a noncompact quantum group,''
   \href{https://arxiv.org/abs/hep-th/9911110}{{\ttfamily hep-th/9911110}}; 
	
	   \bibitem{groenevelt}
  W.~Groenevelt, ``The Wilson function transform," \href{https://arxiv.org/abs/math/0306424}{{\ttfamily arXiv:0306424 [math.CA]}}; ``Wilson function transforms related to Racah coefficients," 	\href{https://arxiv.org/abs/math/0501511}{{\ttfamily arXiv:math/0501511 [math.CA]}}.
	
\bibitem{LeFloch:2017lbt} 
  B.~Le Floch and G.~J.~Turiaci,
  ``AGT/$\mathbb{Z}_2$,''
  JHEP {\bf 1712}, 099 (2017)
     \href{https://arxiv.org/abs/1708.04631}{{\ttfamily [arXiv:1708.04631[hep-th]]}}.
	
\bibitem{Kaplan} 
  H.~Chen, A.~L.~Fitzpatrick, J.~Kaplan, D.~Li and J.~Wang,
  ``Degenerate Operators and the $1/c$ Expansion: Lorentzian Resummations, High Order Computations, and Super-Virasoro Blocks,''
  JHEP {\bf 1703}, 167 (2017)
  \href{https://arxiv.org/abs/1606.02659}{{\ttfamily [arXiv:1606.02659 [hep-th]]}}.   
	
		\bibitem{Gu:2017njx}
  Y.~Gu, A.~Lucas and X.~L.~Qi,
  ``Spread of entanglement in a Sachdev-Ye-Kitaev chain,''
  JHEP {\bf 1709} (2017) 120
  \href{https://arxiv.org/abs/1708.00871}{{\ttfamily [arXiv:1708.00871[hep-th]]}}.
  
	\bibitem{Kourkoulou:2017zaj}
  I.~Kourkoulou and J.~Maldacena,
  ``Pure states in the SYK model and nearly-$AdS_2$ gravity,''
  \href{https://arxiv.org/abs/1707.02325}{{\ttfamily [arXiv:1707.02325 [hep-th]]}}.
	
	\bibitem{Eberlein:2017wah}
  A.~Eberlein, V.~Kasper, S.~Sachdev and J.~Steinberg,
  ``Quantum quench of the Sachdev-Ye-Kitaev Model,''
  Phys.\ Rev.\ B {\bf 96} (2017) no.20,  205123
  \href{https://arxiv.org/abs/1706.07803}{{\ttfamily [arXiv:1706.07803 [cond-mat.str-el]]}}.
	
	\bibitem{Sonner:2017hxc}
  J.~Sonner and M.~Vielma,
  ``Eigenstate thermalization in the Sachdev-Ye-Kitaev model,''
  JHEP {\bf 1711} (2017) 149
  \href{https://arxiv.org/abs/1707.08013}{{\ttfamily [arXiv:1707.08013 [hep-th]]}}.

\bibitem{Seiberg:1990eb} 
  N.~Seiberg,
  ``Notes on quantum Liouville theory and quantum gravity,''
  Prog.\ Theor.\ Phys.\ Suppl.\  {\bf 102}, 319 (1990).
  
\bibitem{Maldacena:2018lmt} 
  J.~Maldacena and X.~L.~Qi,
  ``Eternal traversable wormhole,''
    \href{https://arxiv.org/abs/1804.00491}{{\ttfamily [arXiv:1804.00491 [hep-th]]}}.
  
\bibitem{Haehl:2017pak} 
  F.~M.~Haehl and M.~Rozali,
  ``Fine Grained Chaos in $AdS_2$ Gravity,''
  Phys.\ Rev.\ Lett.\  {\bf 120}, no. 12, 121601 (2018)
  \href{https://arxiv.org/abs/1712.04963}{{\ttfamily [arXiv:1712.04963 [hep-th]]}}.
  
\bibitem{Qi:2018rqm} 
  Y.~H.~Qi, Y.~Seo, S.~J.~Sin and G.~Song,
  ``Schwarzian correction to quantum correlation in SYK model,''
 \href{https://arxiv.org/abs/1804.06164}{{\ttfamily [arXiv:1804.06164[hep-th]]}}.
 
		

  
\end{thebibliography}
\end{document}